\documentclass[aps,pra,notitlepage,superscriptaddress,onecolumn]{revtex4-2}

\usepackage{amsmath, amssymb}
\usepackage{graphicx}	
\usepackage{bm} 
\usepackage{color}
\usepackage{hyperref}
\usepackage[version=4]{mhchem}
\usepackage{physics}
\usepackage{float}

\newcommand{\mr}[1]{\mathrm{#1}}
\newcommand{\mc}[1]{\mathcal{#1}}
\newcommand{\hath}{\hat{H}}
\newcommand{\bmth}{\bm{\theta}}

\begin{document}

\title{Quantum-selected configuration interaction with time-evolved state}

\author{Mathias Mikkelsen}
\email{mathias@qunasys.com}
\author{Yuya O. Nakagawa}
\affiliation{QunaSys Inc., Aqua Hakusan Building 9F, 1-13-7 Hakusan, Bunkyo, Tokyo 113-0001, Japan}

\date{\today}
\begin{abstract} 
Quantum-selected configuration interaction (QSCI) utilizes an input quantum state on a quantum device to select important bases (electron configurations in quantum chemistry) that define a subspace in which to diagonalize a target Hamiltonian, i.e., perform selected configuration interaction, on classical computers.
Previous proposals for preparing a good input state, which is crucial for the quality of QSCI, based on optimization of quantum circuits may suffer from optimization difficulty and require many runs of the quantum device. Here, we propose using a time-evolved state by the target Hamiltonian (for some initial state) as an input of QSCI.
Our proposal is based on the intuition that the time evolution by the Hamiltonian creates electron excitations of various orders when applied to the initial state.
We numerically investigate the accuracy of the energy obtained by the proposed method for quantum chemistry Hamiltonians describing electronic states of small molecules.
Numerical results reveal that our method can yield sufficiently accurate ground-state energies for the investigated molecules. Systematic analysis when increasing the number of qubits in a hydrogen chain shows that the subspace size required for sufficiently accurate results is reasonable at system sizes that cannot be solved by naive classical diagonalization.
Our proposal provides a systematic and optimization-free method to prepare the input state of QSCI and could contribute to practical applications of quantum computers in quantum chemistry calculations.
\end{abstract}

\maketitle 

%%% ------------------------ %%%
\section{Introduction}
\label{sec:introduction}
A major motivation for quantum computers is the simulation of quantum many-body systems~\cite{Feynman1982} and quantum chemistry~\cite{Cao2019,McArdle2020}.
In the early days, the quantum phase estimation (QPE)~\cite{kitaev1995quantum,cleve1998} algorithm was the main focus, but in the last decade quantum-classical hybrid algorithms that can perform calculations even on noisy devices have attracted much attention.
Among these, the variational quantum eigensolver (VQE)~\cite{Peruzzo2014,Tilly2022} has attracted the most attention.
However, its applicability to practical-scale problems of interest is not obvious because of the difficulty of optimization, the so-called barren plateau~\cite{McClean2018Barren}, and the huge number of quantum circuit runs required to perform highly accurate quantum chemistry calculations~\cite{Gonthier2022}.
Therefore, research on utilizing near-term quantum computers with quantum algorithms other than VQE has been active in the past few years, some examples being quantum-enhanced Markov chain Monte Carlo~\cite{Layden2023}, auxiliary-field quantum Monte Carlo with quantum computing of overlaps~\cite{Huggins2022}, the quantum Krylov method \cite{parrish2019quantum,Stair2020} and so on. These methods emphasize more classical processing than VQE, and assume that the ground-state energy is ultimately computed using a classical computer.

One such method, quantum-selected configuration interaction (QSCI), was proposed by a group including one of the authors~\cite{kanno2023quantum}. QSCI calculates the eigenstate energies of a system by performing measurements on an appropriately prepared quantum state in a computational basis, interpreting the obtained measurement results as important electron configurations (basis states in Hilbert space), and diagonalizing the Hamiltonian in the subspace consisting of only those electron configurations with a classical computer.
The difficulty of classical simulation of measurement results for certain quantum states is known~\cite{arute2019quantum}, and QSCI using such quantum states as input theoretically guarantees the possibility of exploring configurations that could not be explored by classical simulation.
Of course, whether the electron configurations output by quantum states are effective in approximating the Hamiltonian energy is a nontrivial question, and research from various perspectives is needed to verify their effectiveness in practice.

One of the problems with QSCI is the lack of a systematic construction of its input state.
The original paper~\cite{kanno2023quantum} proposed the use of VQE, which is less accurate and inherits some of the scalability issues of VQE, while more recently the so-called ADAPT-QSCI method~\cite{nakagawa2023adapt} inspired by ADAPT-VQE~\cite{grimsley2019adaptive} was proposed.
As both of these methods are based on optimization and have overhead, it would be useful to find an optimization-free method to prepare an appropriate input state.
In addition, developing many different state preparation methods allows us to perform QSCI by combining them, potentially leading to a better result than using either method alone.
We note that the large-scale experimental realization of QSCI~\cite{robledomoreno2024} employs an input state based on a classical calculation without optimizing a quantum circuit, but systematic analysis of the input state is unexplored (later we compare a similar ansatz with our proposal in Sec.~\ref{subsec:comparison with CCSD}). 

In this study, we propose to use the time evolution operator acting on a suitable initial state as an input state preparation method for QSCI that does not require optimization. The time-evolution involves a superposition of powers of the Hamiltonian $\hat{H}$ acting on the initial state.
The intuitive reason for adopting this method is that electronically excited states (over the initial state) will be generated by such powers with Born probabilities related to the matrix elements of the Hamiltonian between the initial state and electronically excited states.
The use of the time evolution is similar to the Quantum Krylov method~\cite{parrish2019quantum,Stair2020}, where the time-evolved states themselves define the subspace for classical diagonalization and components of the subspace Hamiltonian are evaluated by quantum computers, but in our proposal the subspace is constructed based on the measurement results of the time-evolved state and components of the subspace Hamiltonian are calculated by classical computers.
We apply our method to the problem of finding the ground state of the electronic Hamiltonian for small molecules, and discuss its energy accuracy, required classical resources and optimal time-evolved states.
We also estimate the number of gates and classical resources needed to run the proposed method on large molecules.

This paper is organized as follows. In Section \ref{sec:ReviewofQSCI} we review the original QSCI proposal introducing the basic framework required for any QSCI-based proposal. In section \ref{sec:proposal} we outline the proposal of this paper which we name Time-Evolved QSCI (TE-QSCI). This includes a practical description of the algorithm and a theoretical motivation. In Section~\ref{sec:numerical results}, we provide a variety of numerical evidence for the utility of TE-QSCI using a variety of smaller molecules and hydrogen chains.
Finally we discuss the prospects of the method and potential future directions in Section~\ref{sec:discsussion}.

%%% ------------------------ %%%
\section{Review of quantum selected configuration interaction (QSCI)}
\label{sec:ReviewofQSCI}
In this section, we present a brief review of QSCI~\cite{kanno2023quantum}. QSCI aims to calculate eigenenergies and eigenstates of a given quantum many-body Hamiltonian by combining quantum and classical computers.
We consider a fermionic Hamiltonian and the corresponding $n_q$-qubit Hamiltonian $\hat{H}$.
The mapping between the fermion and qubit representation is assumed to satisfy that each computational basis state $\ket{\mu} \: (\mu=0,1,\cdots,2^{n_q}-1)$ in the qubit representation corresponds to a single fermionic Fock state.
Most popular mappings such as the Jordan-Wigner~\cite{Jordan1928}, Parity~\cite{Bravyi2002, Seeley2012}, and Bravyi-Kitaev~\cite{Bravyi2002} transformation fulfill this property.
QSCI can be regarded as a selected configuration interaction (CI) method, classical versions of which are often used in quantum chemistry~\cite{helgaker2014molecular, bender1969pr,whitten1969jcp,huron1973jcp,buenker1974tca,buenker1975tca,nakatsuji1983cluster,cimiraglia1987jcc,harrison1991jcp,greer1995jcp,greer1998jcpss, evangelista2014jcp,holmes2016jctc,schriber2016jcp,holmes2016jctc2,tubman2016deterministic,ohtsuka2017jcp,schriber2017jctc,sharma2017semistochastic,chakraborty2018ijqc,coe2018jctc,coe2019jctc,abraham202jctc,tubman2020modern,zhang2020jctc,zhang2021jctc,chilkuri2021jcc,chilkuri2021jctc,goings2021jctc,pineda2021jctc,jeong2021jctc,seth2023jctc}.
In selected CI, we choose a subset of basis states in the Hilbert space and diagonalize the Hamiltonian in the corresponding subspace. QSCI considers a subspace spanned by a set of computational basis states that correspond to Slater determinants of fermions in quantum chemistry applications under our assumption.
We denote the subspace $\mc{S} = \mr{span}\{ \ket{\mu_1}, \cdots, \ket{\mu_R} \}$, where $\mu_k \in \{0,1,\cdots,2^{n_q}-1\}$ for $k=1,\cdots,R$ and $R$ is the dimension of the subspace.
The projected Hamiltonian in the subspace whose size is $R\times R$ is defined by $H^{\mc{S}}_{kl} = \mel{\mu_k}{\hath}{\mu_l}$ and diagonalized to calculate the approximate ground state and its energy using classical computers. If the Hamiltonian is sparse in the sense that it contains a polynomial number of fermion operators or Pauli operators with respect to $n_q$, the matrix component $H^{\mc{S}}_{kl}$ can be computed efficiently by classical computers.
The choice of subspace is vital for obtaining a good approximation of the exact eigenenergies and eigenstates in selected CI. We note that $R$ is assumed to be not so large that we cannot diagonalize the projected Hamiltonian by a classical computer within a reasonable amount of time (see the original proposal of QSCI~\cite{kanno2023quantum} for this point).

QSCI employs a quantum state $\ket{\Phi}$ on a quantum computer to choose the subspace used in selected CI. We prepare $\ket{\Phi}$ on a quantum computer and perform projective measurements on the computational basis. For the state $\ket{\Phi}$ written as
\begin{equation} \label{eq:basis expansion}
\ket{\Phi} = \sum_{\mu=0}^{2^{n_q}-1} \alpha_\mu \ket{\mu},
\end{equation}
an integer $\mu$ is obtained with the probability $|\alpha_\mu|^2$ as a result of the projective measurement.
The projective measurement on a quantum state is sometimes called sampling.
We denote the number of repetitions of the projective measurement by $n_{\text{shots}}$.

The brief outline of the QSCI algorithm is as follows:
\begin{enumerate}
 \item Prepare some input state $\ket{\Phi}$ on a quantum computer and perform the projective measurement on the computational basis $n_{\text{shots}}$ times. The $n_{\text{shots}}$ measurements produces $n_{\text{shots}}$ integers as $\mu_1, \mu_2, \cdots, \mu_{n_{\text{shots}}} \in \{0,1,\cdots,2^{n_q}-1\}$.
 \item Among the results of the projective measurements, choose the $R$ most frequent integers, $\mu_1, \cdots, \mu_R \in \{0,1,\cdots,2^{n_q}-1\}$  by calculating the occurrence frequencies of the integers $\mu$ appearing in the measurement results, defined as $f_\mu = n_\mu/n_{\text{shots}}$, where $n_\mu$ is the number of times $\mu$ is contained in the measurement results. 
 \item Define the subspace $\mc{S} = \mr{span}\{\ket{\mu_1}, \cdots, \ket{\mu_R} \}$ and perform the diagonalization of the projected Hamiltonian onto the subspace $\mc{S}$ using classical computers. This gives the approximate ground state and ground-state energy of the Hamiltonian.
\end{enumerate}

The quality of the approximate energies and eigenstates obtained by QSCI is determined by the choice of the input state $\ket{\Phi}$. The input state must contain important bases to express the exact ground state with large weights so that those bases appear frequently in the measurement results and are picked up in the selected CI calculation.
One of the ideal candidates for such an input state is the exact ground state $\ket{\psi_\mr{GS}}$ of the Hamiltonian $H$ itself because it contains the important bases with large weights almost by definition. Based on this consideration, the original proposal of QSCI~\cite{kanno2023quantum} leveraged VQE with loose optimization to prepare the input state of QSCI in numerical simulations and experiments on quantum hardware.
Later, another study~\cite{nakagawa2023adapt} proposed an adaptive construction of the input state by repetitive executions of QSCI, dubbed ADAPT-QSCI. ADAPT-QSCI enables us to run the whole QSCI algorithm without a heuristic choice of the input state, but there may still be some difficulties in executing it for practical large problems. For example, ADAPT-QSCI requires many repetitions of QSCI.  The precision of the result also depends on the so-called operator pool that must be given by hand. Therefore, it is meaningful to propose another recipe for the input state of QSCI circumventing these difficulties. This study proposes such a method for quantum chemistry Hamiltonians by using the time-evolved state by the target Hamiltonian.

%%% ------------------------ %%%
\section{Our proposal: QSCI with time-evolved state}
\label{sec:proposal}
In this section, we describe our proposal, where the input state of QSCI is generated by applying the time-evolution operator with the target Hamiltonian to a suitable initial state, which we name Time-Evolved QSCI (TE-QSCI). We first present a framework and concrete procedures of the TE-QSCI algorithm with several possible variants and then explain the theoretical motivation for why we expect TE-QSCI to work. We employ the convention $\hbar=1$ throughout this section to ease the notation. 

\subsection{Algorithm description}
In this study, we specifically consider quantum chemistry problems for electronic states of molecules, described by the Hamiltonian
\begin{equation} \label{eq:Hamiltonian}
 \hat{H}_F = \sum_{\sigma=\uparrow, \downarrow} \sum_{p,q=1}^{N_O} h_{pq} \hat{c}_{p,\sigma}^\dag \hat{c}_{q,\sigma} + \frac{1}{2} \sum_{\sigma,\tau=\uparrow, \downarrow} \sum_{p,q,s,r=1}^{N_O} V_{pqrs} \hat{c}_{p,\sigma}^\dag \hat{c}_{q,\tau}^\dag \hat{c}_{r,\tau} \hat{c}_{s,\sigma},
\end{equation}
where $\hat{c}_{p,\sigma}^\dag (\hat{c}_{p,\sigma})$ is a creation (annihilation) operator of electrons in the orbital $p$ with spin $\sigma$, $N_O$ is the number of molecular orbitals and $h_{pq}, V_{pqrs}$ are electron integrals computed efficiently by classical computers.
When this Hamiltonian acts on some reference state $\ket{\psi}$ as $\hath \ket{\psi}$, the first term ($\hat{c}^\dag \hat{c}$) generates single-electron excitations over $\ket{\psi}$ and the second term ($\hat{c}^\dag \hat{c}^\dag \hat{c} \hat{c}$) generates double-electron excitations. Note that our proposal is not limited to this type of Hamiltonian as long as the system allows the notion of particle excitations on some reference state and the Hamiltonian is written in operators describing such excitations.

Our proposal to find the approximate ground-state and ground state energy by using classical and quantum computers, leverages the time-evolved state from a given initial state $\ket{\psi_I}$ for some time $t$, 
\begin{equation}
\ket{\psi(t)} \equiv e^{-i\hath t}\ket{\psi_I},
\end{equation}
as the input of QSCI.
The outline of the TE-QSCI algorithm is as follows:
\begin{enumerate} 
    \item 
    Prepare the initial state $\ket{\psi_I}$ on a quantum computer and choose times $t_1, \cdots, t_K$ ($K$ can be any positive integer).
    \item 
     For each $k=1,\cdots,K$, apply the time-evolution operator $e^{-i\hath t_k}$ to the initial state and obtain the time-evolved state $\ket{\psi(t_k)}=e^{-i\hath t_k}\ket{\psi_I}$.
    \item
     Perform the projective measurement (sampling) on the state $\ket{\psi(t_k)}$ $N_k$ times, which yields the result (a set of integers) $\mu_1^{(k)}, \cdots, \mu_{N_k}^{(k)} \in \{0,1,\cdots,2^{n_q}-1\}$.
     \item 
     Collect all of the measurement results for $k=1,\cdots, K$ and post-process them by classical computers to define the subspace for the CI calculation (exact diagonalization). This part may take various forms and we explain two representative cases below.
     \item Calculate the approximate ground-state energy and ground state by performing the CI calculation in the subspace by classical computers.
\end{enumerate}
TE-QSCI can take various forms in the actual implementation depending on (1) the choice of the initial state $\ket{\psi_I}$ (step 1), (2) the choice of the time $t_1, ..., t_K$ (step 1), and (3) the procedure to define the subspace (step 4).
We explain two representative variants of TE-QSCI here, \textit{single-time TE-QSCI} and \textit{time-average TE-QSCI}.

\paragraph*{\bf{Single-time TE-QSCI:}}
Single-time TE-QSCI is the simplest TE-QSCI implementation.
The number of the time-evolved states is $K=1$, which means that we perform the original QSCI by taking the input state $\ket{\psi(t_1)}$ as input.
The subspace for the CI calculation is determined by the measurement results on the single state $\ket{\psi(t_1)}$, as reviewed in Sec.~\ref{sec:ReviewofQSCI}.
We consider single-time TE-QSCI in most of the numerical investigations in the next section because of its simpleness, and we also discuss the appropriate time $t_1$ to obtain an accurate energy and eigenstate.

\paragraph*{\bf{Time-average TE-QSCI:}}
Time-average QSCI is another implementation of TE-QSCI.
We take $K>1$ in this case and set the number of shots for measurements on each time-evolved state $\ket{\psi(t_k)} (k=1,\cdots,K)$ to the same value $N_s$ (step 3).
To construct the subspace in step 4, we simply concatenate all the measurement results (observed integers) for $k=1,\cdots,K$ as $\{\mu_{i_1}^{(1)}\}_{i_1=1}^{N_s} \cup \cdots \cup  \{\mu_{i_K}^{(K)}\}_{i_K=1}^{N_s}$ and then pick up the $R$ most frequent integers.
The name of \textit{time-average} is derived after the following argument: 
When we choose the times as $t_k = T_1+k \frac{T_2-T_1}{K} \: (k=1,\cdots,K)$ for some times $T_1,T_2$, in the limit of $K\to\infty$, the construction of the subspace in time-average TE-QSCI corresponds to that of the original QSCI with inputting the time-averaged (mixed) state $\frac{1}{T_2-T_1}\int_{T_1}^{T_2} dt \ketbra{\psi(t)}$.
Note that QSCI with multiple input states was already discussed in the original proposal of QSCI~\cite{kanno2023quantum} to calculate the excited states of the Hamiltonian.

We comment on some possible advantages of TE-QSCI. First, compared with other preparation methods for the input state of QSCI mentioned in Sec.~\ref{sec:introduction}, TE-QSCI does not need optimization of the quantum circuit. This feature not only circumvents the difficulties associated with the circuit optimization that is often encountered in quantum-classical hybrid algorithms~\cite{McClean2018Barren} but also reduces the number of runs on the quantum computer.
For example, in ADAPT-QSCI~\cite{nakagawa2023adapt}, the input state of QSCI is adaptively constructed by repeating the execution of QSCI many times (e.g., 50 times for the 12 qubit \ce{H6} molecule), and the number of repetitions can grow with the system size.
Second, the time-evolution operator $e^{-i\hath t}$ does not need to be so precise in TE-QSCI compared with other quantum algorithms such as quantum phase estimation and the quantum Krylov method. This is because TE-QSCI only uses the time-evolved state $\ket{\psi(t)} = e^{-i\hath t} \ket{\psi_I}$ to pick up the bases of the subspace and never evaluates the expectation value of the state as in the quantum Krylov method which is more noise-sensitive.
Moreover, we do not need the controlled time evolution or any ancillary qubits as is required for quantum phase estimation.

\subsection{Theoretical background and motivation for TE-QSCI \label{sec:theoretical motivation}}
Intuitively, the reason why we expect TE-QSCI to work is as follows. The time-evolved state is formally written as
\begin{equation}
 \ket{\psi(t)} = e^{-i\hath t}\ket{\psi_I}
 = \ket{\psi_I} - i \hath t \ket{\psi_I} + \frac{(-i\hath t)^2}{2} \ket{\psi_I} + \cdots.
\end{equation}
Since the Hamiltonian $\hath$ [Eq.~\eqref{eq:Hamiltonian}] contains the terms of up to second-order excitations of electrons, the $k$-th order term $\frac{(-i\hath t)^k}{k!}\ket{\psi_I}$ in the above expression include the excitations up to $2k$-th order on the initial state $\ket{\psi_I}$.
Therefore, it is naively expected that the time-evolved state $\ket{\psi(t)}$ possesses the various important (Fock) states excited over the initial state.
These states are naturally candidates for selected CI calculations in quantum chemistry, similar to configuration interaction singles and doubles (CISD) when taking the initial state as the Hartree-Fock state. This mechanism could make the time-evolved state a nice candidate for the input state of QSCI.

To refine the intuitive argument above, let us consider the mathematical expression of the probability of measuring the computational basis state $\ket{\mu}$ for the time-evolved state $\ket{\psi(t)}$, defined as $P_\mu(t)= \abs{\braket{\mu}{\psi(t)}}^2$. 
In Appendix~\ref{app:seriesexpansion} we show that $P_\mu(t)$ can be written as 
\begin{align}
P_{\mu}(t) & = |b_{\mu}|^2 +\sum_{k=1}^{\infty} \left(2 t^{k} \frac{\text{Re}\left[ (i)^{k} b_{\mu} \mel{\psi_{I}}{\hat{H}^{k}}{\mu} \right] }{k!} + t^{2k} \frac{ \abs{\mel{\mu}{\hat{H}^{k}}{\psi_{I}}}^2}{(k!)^2}  \nonumber \right. \\
+ & \left. 2\sum_{k'=1}^{\infty} t^{2k+k'} \frac{ \text{Re}\left[(i)^{k'} \mel{\mu}{\hat{H}^{k}}{\psi_{I}}  \mel{\psi_{I}}{\hat{H}^{k+k'}}{\mu} \right] }{k! (k+k')!} \right),
\label{eq:probabilityschematicseriesexpansion}    
\end{align}
where $b_{\mu}=\braket{\mu}{\psi_I}$ is the overlap between $\ket{\mu}$ and the initial state.
The probabilities for measuring the states $\mu$ with $b_{\mu}=0$, i.e., the states not included in the initial state at all, grow algebraically as $t^{2k}$ for small $t$, where $k$ is determined by how the power of the Hamiltonian connects these states with the initial state, $\mel{\mu}{\hat{H}^{k}}{\psi_{I}}$.
In other words, the probability $P_\mu(t)$ is related to the $k$-th order perturbation of the Hamiltonian applied to the initial state.
Note the similarity to classical selected CI algorithms like \cite{holmes2016jctc2}, for which states are iteratively chosen based on the value of $\mel{\mu}{\hat{H}}{\psi^{(k)}}$, where $\ket{\psi^{(k)}}$ is the ground state of the Hamiltonian in the subspace of the previous step. In contrast, TE-QSCI samples multiple higher-order excitations simultaneously in a single step.
$P_\mu(t)$ for the state $\mu$ that can be generated from $\ket{\psi_I}$ by the $k$-th order perturbation scales as $t^{2k}$ for small $t$, which means that we need to take not-so-small $t$ to sample the higher-order excitations over the initial state. We confirm this scaling numerically in Sec.~\ref{subsec:time evolution of P_mu}. 

The series expansion in Eq.~\eqref{eq:probabilityschematicseriesexpansion} is useful mostly for small $t$. When we denote the eigenstates and eigenenergies of the Hamiltonian as $\ket{\psi_n}$ and $E_n$, respectively, the general expression for $P_\mu(t)$ is
\begin{align}
P_\mu(t)&= \left| \sum_{n}e^{-i E_n t} c_n^I c_n^\mu \right|^2 \nonumber \\
&= \sum_{n} |c_n^I|^2 |c_n^\mu|^2 + 2\sum_{n_1 < n_2} \cos((E_{n_2}-E_{n_1})t) \text{Re}\left(c_{n_1}^I (c_{n_2}^I)^* c_{n_1}^\mu (c_{n_2}^\mu)^* \right) 
\label{eq:timedependentprobabilities}
\end{align}
where $c_n^I= \braket{\psi_n}{\psi_I}$ and $c_n^\mu = \braket{\mu}{\psi_n}$.
Although it is difficult to extract any information from this expression for general $t$ because of its complicated interference of terms, we can easily see that the infinite time-average of the probabilities is given by
\begin{equation}
\bar{P}_{\mu} = \lim_{T\rightarrow \infty } \frac{1}{T}\int_0^T  P_{\mu}(t) dt = \sum_{n} |c_n^I|^2 |c_n^\mu|^2.
\label{eq:infinitetimeaverage}
\end{equation}
From this expression, one notices that QSCI based on the exact ground-state input can be mimicked if $|c_0^I|^2 = \abs{\braket{\psi_0}{\psi_I}}^2$ is dominant.
As discussed in the original proposal of QSCI, the exact ground-state is one of the ideal input states of QSCI because it contains the important bases for the CI calculation by definition.
Therefore, we may argue that time-average QSCI with a sufficiently large time $T$ with a proper initial state whose overlap with the exact ground state is large performs similarly to QSCI based on the exact ground state, resulting in a precise approximation of the ground state and energy.
Note that this is not necessarily true for directly measuring an initial state with a large ground-state fidelity $|c_0^I|^2$ as interference due to the second sum in Eq.~\eqref{eq:timedependentprobabilities} (cosine evaluates to 1 for $t=0$) can change the effective distribution of the measurement result, which means that the ground state is not sampled proportionally to $|c_0^I|^2$.
This is clear for the Hartree-Fock state, which only contains a single computational state, despite having a relatively large overlap with the ground state in many cases.
Sampling based on the time-evolved state is different than that on the initial state despite $\abs{\braket{\psi_0}{\psi_I}}^2 = \abs{\braket{\psi_0}{\psi(t)}}^2$ during the time evolution.

%%% ------------------------ %%%
\section{Numerical results \label{sec:numerical results}}
In this section, we investigate the performance of TE-QSCI numerically for electronic Hamiltonians describing small molecules. 
Sections~\ref{subsec:time evolution of P_mu}-\ref{subsec:time-average TE-QSCI} consider the Hartree-Fock state as the initial state of TE-QSCI.
Before investigating TE-QSCI directly, in Sec.~\ref{subsec:time evolution of P_mu} we numerically investigate the time evolution of $P_{\mu}(t)$ to validate the analytical formula presented in the previous section. In Sec.~\ref{subsec:performance of single TE QSCI}, we study how the accuracy of single-time TE-QSCI depends on the parameters such as the time $t$, the Trotter step size used in implementing the time-evolution operator, the dimension of the subspace for the CI calculation, and the number of shots.
Then, we compare the performance of single-time QSCI at the optimal time $t$ with that of time-average TE-QSCI in Sec.~\ref{subsec:time-average TE-QSCI}.
In Sec.~\ref{subsec:comparison with CCSD}, we compare the accuracy of TE-QSCI with the Hartree-Fock initial state to a different initial state, namely, the unitary coupled cluster singles and doubles (unitary CCSD, or UCCSD) ansatz whose parameters are determined by a classical CCSD calculation. In addition to using this as an initial state for time evolution, we also compare QSCI directly sampling the UCCSD ansatz. Section~\ref{subsec:scaling analysis} presents how the classical and quantum resources for TE-QSCI (using both the Hartree-Fock and UCCSD input states) scale with system size. 

The common details of the numerical calculations in this section are as follows. We employ the atomic unit throughout this section, so the unit of time is $\hbar/\mr{Hartree} \sim 2.42 \times 10^{-17}$ s. We consider hydrogen chain molecules where the hydrogen atoms are aligned in a straight line with a distance of $1$ \AA. We also consider \ce{N2} and \ce{NH3} molecules whose atomic coordinates are taken from the CCCBDB database~\cite{NIST_CCCBDB} as the most stable structure at the level of Hartree-Fock/STO-3G. We construct the fermionic Hamiltonian on the form of Eq.~\eqref{eq:Hamiltonian} using the Hartree-Fock orbitals with the STO-3G basis set, calculated by the numerical package PySCF~\cite{Sun2018,Sun2020}.
We construct two types of Hamiltonian for the \ce{N2} molecule: One adopts the active space approximation consisting of ten electrons with eight orbitals (8o,10e) around highest occupied
molecular orbital (HOMO) and lowest occupied
molecular orbital (LUMO) and the other considers the full space of 14 electrons with 10 orbitals (10o,14e). The Jordan-Wigner transformation implemented in OpenFermion~\cite{McClean2020} is then applied to obtain the qubit representation of the Hamiltonians. In order to understand the performance of TE-QSCI, it is useful to compare with QSCI based on sampling (inputting) the exact ground state as the reference of the quality of the obtained energy, which was largely considered in~Ref.~\cite{kanno2023quantum}. We will refer to QSCI with inputting the exact ground state as \textit{GS-QSCI}. Note that GS-QSCI is not an actual protocol in practice, as it assumes the exact ground state is already known.
The number of qubits $n_q$ for the Hamiltonian, the size of the relevant ground-state Hilbert space sector $D$, and the dimension $R_{GS}$ of the subspace required to obtain sufficiently accurate results for GS-QSCI, which is defined such that taking the $R_{GS}$ bases with the largest amplitudes of the exact ground state generates the energy within the error of $10^{-3}$ Hartree compared with the exact one, are summarized in Table~\ref{tab:basicmoleculeinfo}.

In some parts of the numerical simulations, the time-evolution operator $e^{-i\hat{H}t}$ is approximated by the first-order Trotter expansion; for the qubit Hamiltonian written as  
$\hat{H}= \sum_{j} w_j \hat{P}_j$, where $w_j$ is a real coefficient and $\hat{P}_j$ is a Pauli operator, the time-evolution operator is implemented as 
\begin{align}
e^{-i\sum_{j} w_j \hat{P}_j t}  \approx \left(\prod_j e^{-i w_j \hat{P}_j \Delta t}\right)^{n_\mr{step}},
\end{align}
where $n_\mr{step}$ is the number of Trotter steps and $\Delta t= t/n_\mr{step}$ is the Trotter step size. The quantum circuit simulation is performed with Qulacs~\cite{Suzuki2021} without assuming any noise in the circuit. 

\begin{table*}[htb!]
\begin{tabular}{|c|c|c|c|c|c|c|c|}
\hline
Molecule & \ce{H6} & \ce{H8}  & \ce{H10} & \ce{N2}(8o,10e)  & \ce{N2}(10o,14e)  & \ce{NH3}  \\ \hline
$n_q$ & 12 & 16  & 20  & 16  & 20  & 16 \\ \hline
$D$ & 400 & 4900  & 63504  & 3136  & 14400  & 3136 \\ \hline
$R_{GS}$ & 85  & 685  & 4834  & 116 & 124 & 76 \\ \hline
\end{tabular}
\caption{Summary of the molecules investigated in this section. $n_q$ is the number of qubits required to describe the molecule. 
$D$ is the Hilbert space dimension of the sector containing the ground state, considering the number of electrons and restricting the total $z$ component of the spin to zero.
$R_{GS}$ is the number of bases (the dimension of the subspace) required to obtain the energy with error smaller than $10^{-3}$ Hartree with respect to the exact ground-state energy using GS-QSCI.}
\label{tab:basicmoleculeinfo}
\end{table*}

%%% ------ %%%
\subsection{Time evolution $P_{\mu}(t)$ of the probability to measure the computational basis state $\ket{\mu}$ \label{subsec:time evolution of P_mu} }
\begin{figure*}[htb]
\includegraphics[width=0.6\linewidth]{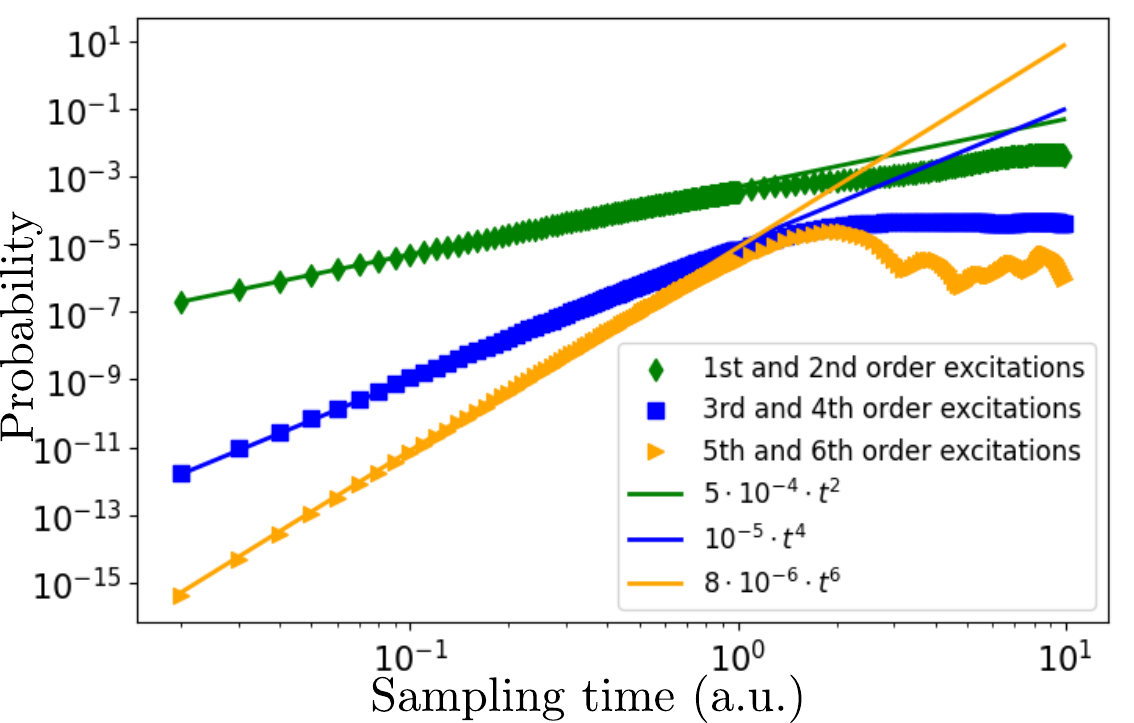}
\caption{Average of the probability of measuring the computational basis state $\ket{\mu}$ for the time-evolved state, $P_\mu(t)=\abs{\braket{\mu}{\psi(t)}}^2$, among the $R=850$ states with largest probability at $t= 1$ for the \ce{H8} molecule. The average is taken for first- and second-order, third- and fourth-order, and fifth- and sixth-order electron excitations.}
\label{fig:Hydrogenexcitationsandprobabilities} 
\end{figure*}

As a test bed for investigating the time evolution of $P_\mu(t)$ in Eq.~\eqref{eq:probabilityschematicseriesexpansion}, we consider the \ce{H8} molecule.
Since we take the Hartree-Fock state as the initial state $\ket{\psi_I}$, the notion of the order of electron excitations is naturally associated with each computational basis state $\ket{\mu}$.
Because $b_\mu=0$ holds in Eq.~\eqref{eq:probabilityschematicseriesexpansion} for states other than the Hartree-Fock state itself, $P_\mu(t)$ grows as $t^{2k}$ for small $t$, where $k$ is the smallest integer satisfying $\mel{\mu}{\hath^k}{\psi_I} \neq 0$.
More simply speaking, $P_\mu(t)$ grows as $t^{l'}$ for small $t$ when $\ket{\mu}$ is the $l$-th order electron excitation (over the Hartree-Fock state), where $l' = l$ ($l+1$) for even (odd) $l$.

Figure~\ref{fig:Hydrogenexcitationsandprobabilities} shows the average of $P_{\mu}(t)$ for one/two-, three/four- and five/six-electron excitation states within the computational basis states having the $R=850$ largest probabilities at $t=1$. The exact time evolution is considered in this figure. For small $t < 1$, these obey the expected $t^{2}$,$t^{4}$, and $t^{6}$ scaling, respectively.
Deviation from this starts happening at around $t\sim 1$, where the series expansion in Eq.~\eqref{eq:probabilityschematicseriesexpansion} is not so meaningful and many orders of the perturbations start to contribute to the probability $P_\mu(t)$.

%%% ------ %%%
\subsection{Performance of single-time TE-QSCI based on the Hartree-Fock state: exact and Trotterized time evolution}
\label{subsec:performance of single TE QSCI}
\begin{figure*}[htb]
\includegraphics[width=1\linewidth]{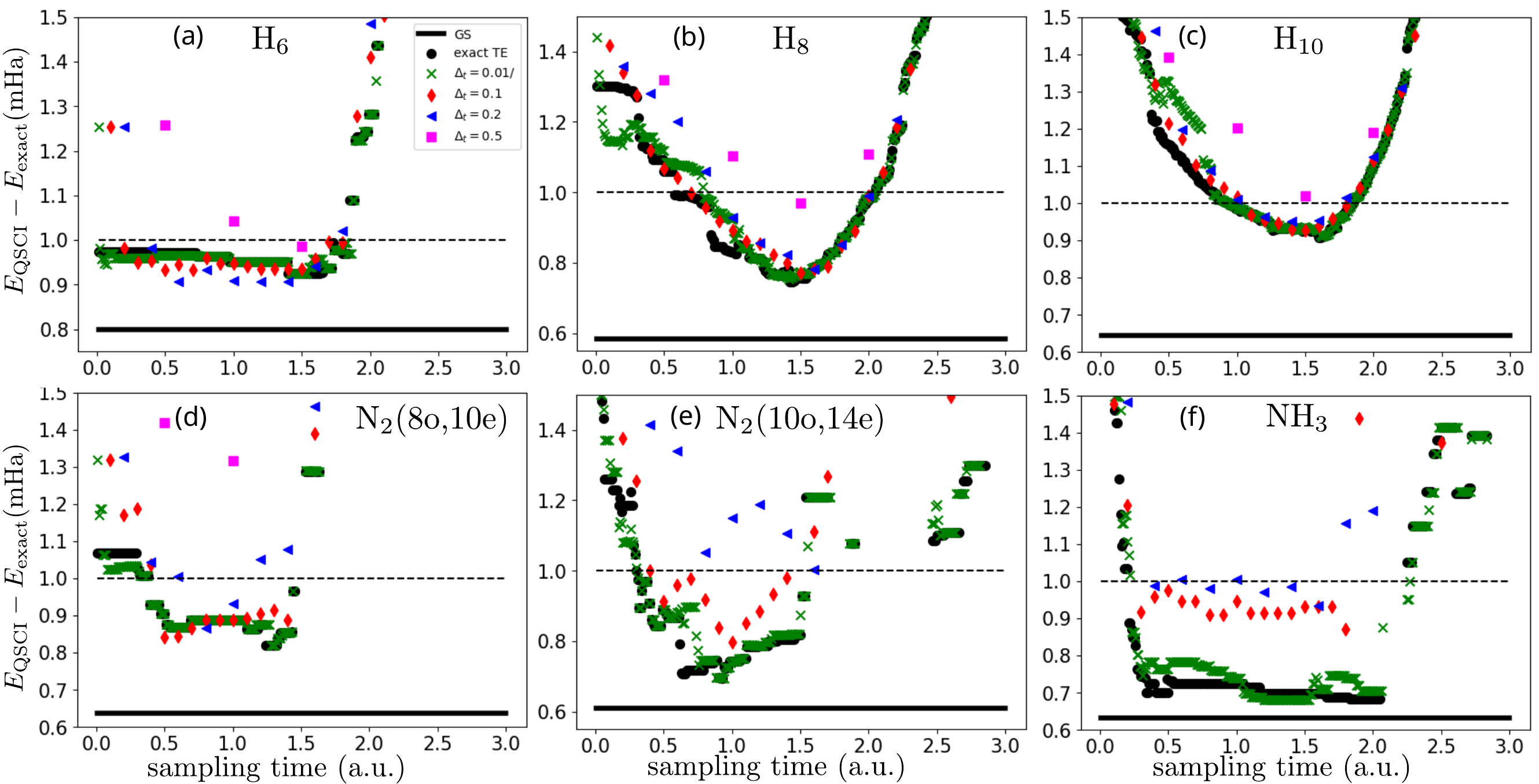}
\caption{
Difference between the exact ground-state energy and the energy obtained by single-time QSCI with various Trotter step sizes $\Delta t$ for the approximation of the time-evolution operator. The solid black line corresponds to GS-QSCI, while the dashed black line corresponds to $10^{-3}$ Hartree. The black circles represent the results of the exact time evolution. The dimension of the subspace in TE-QSCI is $R=90, 850, 6000, 130, 160$ and $100$ for panels (a)-(f), respectively.}
\label{fig:timedependentqsci} 
\end{figure*}

Here, we investigate the performance of single-time TE-QSCI with the Hartree-Fock initial state in detail from several points of view. Figure~\ref{fig:timedependentqsci} shows the difference between the exact ground-state energy and the approximate energy obtained by single-time TE-QSCI as a function of the sampling time $t$ of the time-evolved state $\ket{\psi(t)}$.
We consider both exact time evolution and first-order Trotterization of the time-evolution operators with various step sizes $\Delta t$. These results are based on the so-called state vector simulation of the quantum circuit where the effects of noise and sampling with a finite number of shots are not considered, so the $R$ computational basis states (or electron configurations) with largest amplitudes in $\ket{\psi(t)}$ are used to construct the subspace for the CI calculation. The dimension of the subspace $R$ is taken slightly larger than $R_{GS}$ but much smaller than the dimension $D$ of the relevant sector of the Hilbert space.
From this figure, we see that single-time QSCI can achieve sufficiently accurate results whose errors are below $10^{-3}$ Hartree depending on the value of $t$.
A smaller Trotter step (or more accurate Trotterization of the time-evolution operator) results in a more accurate energy in general. The results also imply that an optimal range for the sampling time exists where the obtained energies are most accurate. Interestingly, the optimal time range is within $t \in [0.5,2.0]$ for all the molecules studied here, despite large differences in the energy itself (e.g., the exact ground-state energy is $-3.2361$ Hartree for \ce{H6} and $-55.5074$ Hartree for \ce{NH3}).

\begin{figure*}[htb!]
\includegraphics[width=1\linewidth]{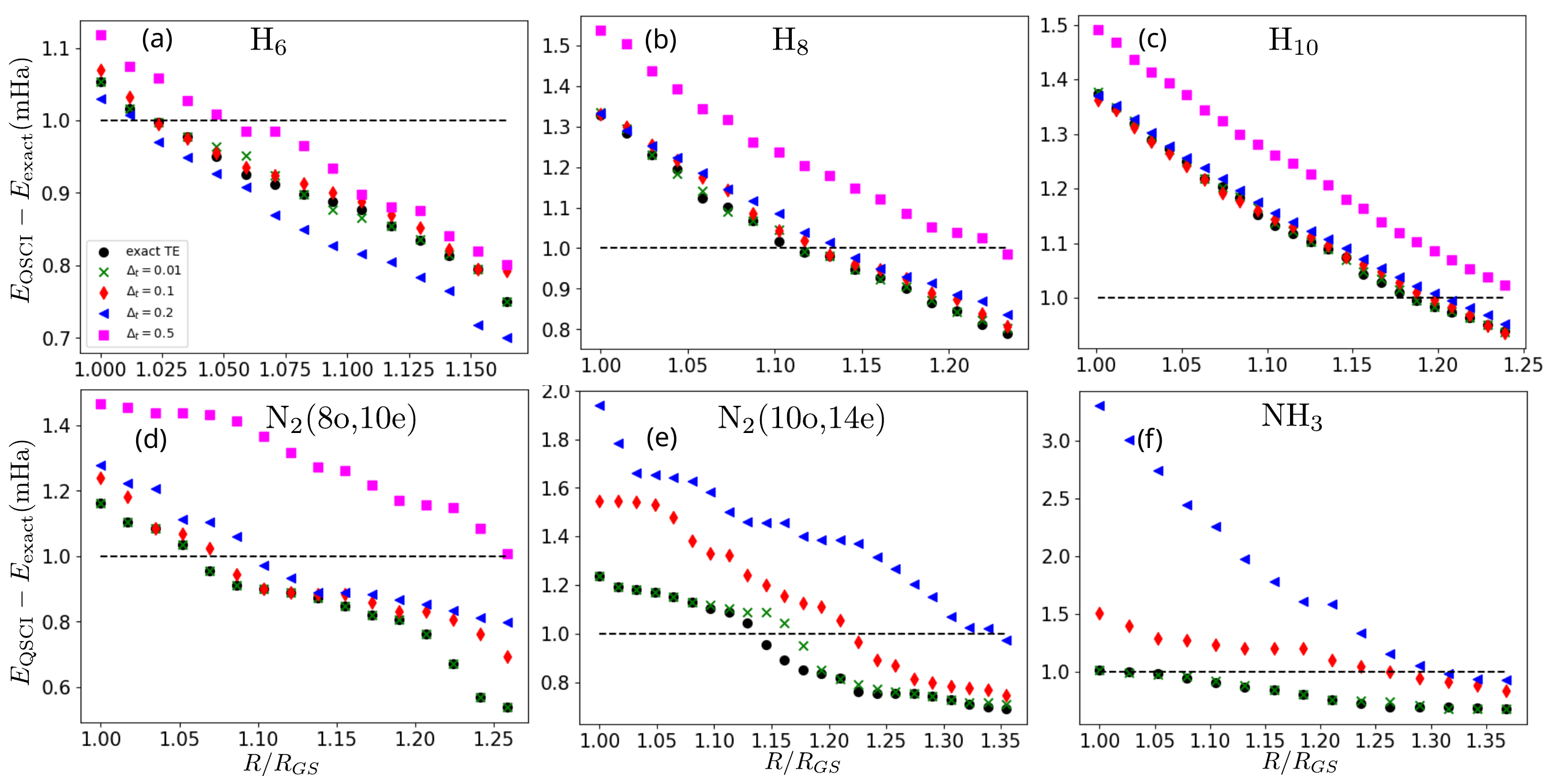}
\caption{
Difference between the exact ground-state energy and the energy obtained by single-time TE-QSCI with the Hartree-Fock initial state as a function $R/R_{GS}$ at fixed times for different molecules. The striped black line corresponds to $10^{-3}$ Hartree and the black circles correspond to the results of the exact time evolution. The sampling times for panels (a)-(c) are all $t=1.4$ except for $\Delta t = 0.5$, where we set $t=1.5$. The sampling times for panels (d),(e), and (f) are $t=1, 1,$ and $1.4$, respectively.
\label{fig:Rdependentqsci}
}
\end{figure*}

Next, we analyze the required dimension of the subspace $R$ in the CI calculation of single-time TE-QSCI. The reference value for $R$ is that for GS-QSCI, $R_{GS}$, summarized in Table~\ref{tab:basicmoleculeinfo} at the beginning of this section and it allows us to evaluate the results of different molecules with a unified criterion. We present the energy difference between the exact energy and output of single-time TE-QSCI as a function of $R/R_{GS}$ in Fig.~\ref{fig:Rdependentqsci}, fixing the time at the optimal value.
In general, TE-QSCI requires a larger $R$ than GS-QSCI but it also depends on the specific molecule; for example, it is rather small for \ce{NH3} with small Trotter steps compared to the other 16 qubit examples.
This is partially explained by the Hartree-Fock state of \ce{NH3} having a large fidelity to the ground state $|\langle HF| \psi_{GS}\rangle|^2=0.970245$. A more systematic investigation of the fidelity of the initial state to the exact ground state is presented in Appendix~\ref{app:fidelitydependence}.
A second important observation in Fig.~\ref{fig:Rdependentqsci} is that an accurate energy below the error of $10^{-3}$ Hartree can still be obtained for large Trotter steps (such as $\Delta t = 0.2$, which is large considering the evolution time is $t\sim 1$), as long as we use a larger $R$ and pay the higher cost of the classical computation in the CI calculation.
Note that \ce{N2} and \ce{NH3} molecules are more sensitive to the Trotter error than the hydrogen chains. This is possibly because of the smaller $R_{GS}$ (and therefore $R$ in the plot) for these molecules, while the larger subspace required for the hydrogen chains can host the unnecessary bases due to the Trotter error and still give an accurate result. 
A more detailed investigation of the Trotter error in terms of the fidelity and energy conservation is presented in Appendix~\ref{app:trottererror}.

\begin{figure*}[htb!]
\includegraphics[width=1\linewidth]{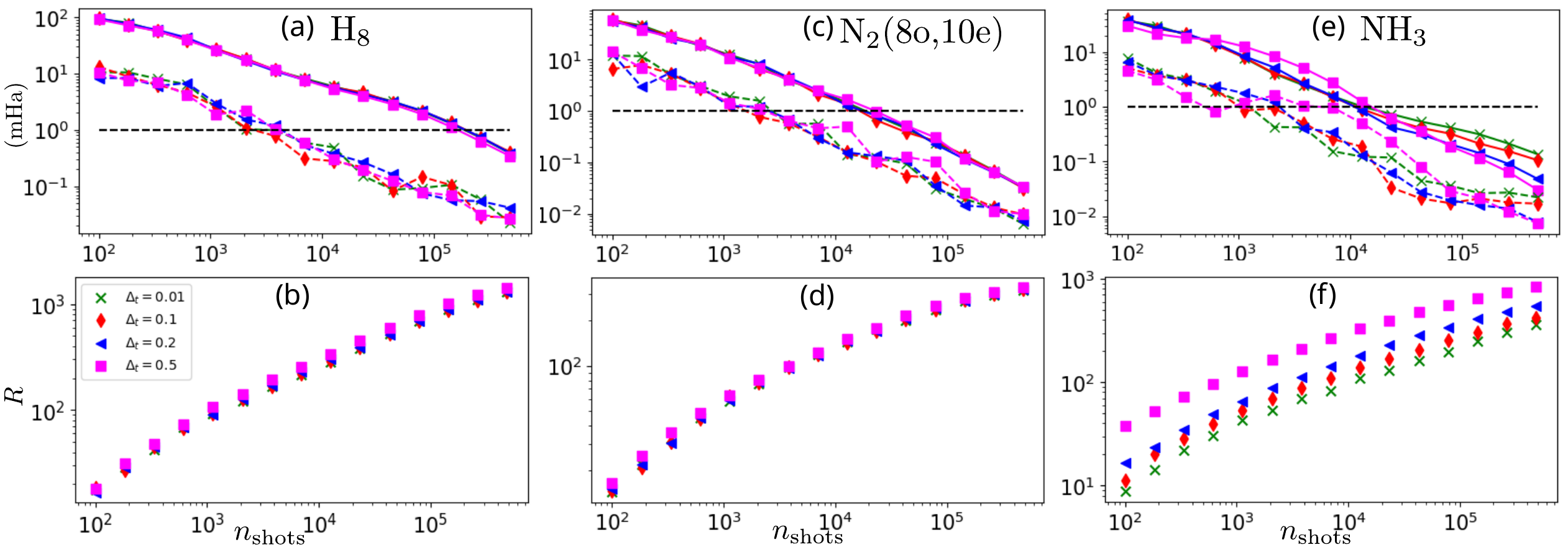}
\caption{
The results of single-time TE-QSCI with the Hartree-Fock initial state considering the effect of the fluctuation of the measurement result with a finite number of shots $n_{\text{shots}}$ at a fixed time $t$.
The dimension of the subspace $R$ is chosen as the number of all distinct bases (electron configurations) obtained by the sampling.
(a), (c), (e): Difference between the exact ground-state energy and the energy obtained by single-time QSCI with various Trotter step size $\Delta t$. We run ten simulations and the mean of the difference is plotted by the markers connected with a full line, while the standard deviation of the obtained energy is plotted by the markers connected by the striped line. The striped black line corresponds to $10^{-3}$ Hartree. 
(b), (d), (f): The average value of $R$ for ten simulations. 
The sampling time $t$ is the same as in Fig.~\ref{fig:Rdependentqsci} except for \ce{NH3} with $\Delta t=0.5$, where we set $t=1.5$.
\label{fig:samplingdependentqsci} }
\end{figure*}

So far, we have investigated single-time TE-QSCI based on the exact state vector simulation, but on real quantum hardware the time-evolved state will be sampled with a finite number of measurements.
The measurements with a finite number of shots introduce the fluctuation of the measurement results of computational basis state, so the resulting output of TE-QSCI also fluctuates.
In Fig.~\ref{fig:samplingdependentqsci} we investigate the effect of a finite number of shots for a fixed $t$ at various values of the Trotter step size $\Delta t$ for the \ce{H8}, \ce{N2}(8o,10e), and \ce{NH3} molecules.
Contrary to the two previous figures in this subsection, the dimension of the subspace $R$ in this figure represents the number of all distinct computational bases (electron configurations) among $n_{\text{shots}}$ measurement results.
We run the simulation ten times to determine the sensitivity to the fluctuations of the measurement results, plotting the mean of the energy error to the exact energy as well as the sample standard deviation of the obtained energies. 
One can see from Fig.~\ref{fig:samplingdependentqsci} that accurate energy is achievable even when considering the shot fluctuation of the measurement results with the standard deviation of the obtained energy being smaller than its mean by about an order of magnitude. 
The energy of single-time TE-QSCI is almost independent of the Trotter step size in this case, where we take all distinct electron configurations appearing in the measurement results to construct the subspace. This treatment results in larger $R$ at the same number of shots for the cases with larger Trotter steps since the larger Trotter error in the time-evolution operator tends to generate a more ``messy" state having a lot of small nonzero amplitudes.
The large $R$ generally yields an accurate energy, compensating for the Trotter error with the drawback of a higher computational cost of the classical diagonalization.

Figures~\ref{fig:Rdependentqsci} and \ref{fig:samplingdependentqsci} suggest that it is possible to trade quantum resources for classical resources, i.e., even when using a less accurate Trotter approximation requiring a shallower circuit, the same accuracy can be obtained with the same number of measurements at the cost of having to diagonalize a larger matrix classically. This trade off is one of the interesting features of TE-QSCI although it is not desirable to push too far in this direction as the classical diagonalization is a major bottleneck for maximal feasible system sizes. 

%%% ----- %%%
\subsection{Performance of time-average TE-QSCI \label{subsec:time-average TE-QSCI}}
\begin{table*}[htb!]
\begin{tabular}{|l|l|l|l|l|}
\hline
Molecule  & \ce{H6} &  \ce{H8} & \ce{N2}(8o,10e)  & \ce{NH3}   \\ \hline
$E_{\text{QSCI}}-E_{\text{exact}}$ at optimal time [mHa] & 0.93  & 0.78  &  0.86 &  0.70 \\ \hline
$E_{\text{QSCI}}-E_{\text{exact}}$ for infinite time average [mHa] & 2.01 & 1.78 & 2.86 & 1.34 \\ \hline
$R$  & 90 & 850 & 130  &  100 \\ \hline
\end{tabular}
\caption{Performance of time-average TE-QSCI with the Hartree-Fock initial state when we take the infinite time average in Eq.~\eqref{eq:infinitetimeaverage} compared with single-time TE-QSCI at the optimal time.
The optimal time values are $t=1.4$ for \ce{H6},\ce{H8} and \ce{NH3}, $t=1.0$ for \ce{N2}(8o,10e). $R$ is the dimension of the subspace for the CI calculation.}
\label{tab:infinitetimeaveragecomparison}
\end{table*}

\begin{table}[htb!]
\begin{tabular}{|l|l|l|l|l|l|l|}
\hline
time [a.u.] &$t=1.4$ &  $t\in[1.0,2.0]$ & $t\in[0.5,1.5]$  & $t\in[1.0,2.5]$ & $t\in[0.5,2.5]$ \\ \hline
$E_\mr{TE-QSCI} - E_\mr{exact}$ [mHa]  & 0.93   & 0.92 & 1.06  &  1.00 &  1.01  \\ \hline
\end{tabular}
\caption{Performance of time-average TE-QSCI with the Hartree-Fock initial state for various time ranges for \ce{H8}, based on numerical simulation using the Trotterized time-evolution operator with $\Delta t = 0.1$ including the effect of a finite number of shots (the results of ten simulations are averaged).
Other details are described in the main text.
\label{tab:multipletimecomparison} }
\end{table}

We investigate the performance of time-average QSCI in this subsection.
First, in Table~\ref{tab:infinitetimeaveragecomparison}, we compare the accuracy of the energy obtained for the infinite-time average calculated according to Eq.~\eqref{eq:infinitetimeaverage} with that of single-time TE-QSCI with the Hartree-Fock initial state at the optimal time.
The time-evolution operator is simulated exactly without Trotterization, and the $R$ computational basis states (electron configurations) with the largest amplitudes are picked up to construct the subspace. 
The results in Table~\ref{tab:infinitetimeaveragecomparison} indicate that the infinite-time-average TE-QSCI is not as accurate as single-time QSCI at the optimal time.
GS-QSCI using the same $R$ can reach the precision of $10^{-3}$ Hartree, but the infinite-time-average TE-QSCI is not equivalent to it because the overlap between the initial state and the exact ground state is not unity and the effect of other excited states is present.

However, we also find that time-average TE-QSCI can achieve accuracy competitive with single-time TE-QSCI at the optimal time depending on the choice of the time range for averaging.
Table~\ref{tab:multipletimecomparison} presents the result of time-average TE-QSCI for various time ranges including the optimal time for the \ce{H8} molecule.
These results are based on the Trotterized time-evolution operator with $\Delta t=0.1$.
The time average is taken for $t \in [T_0, T_1]$ by discretizing the range with $\delta t=0.1$. 
Concretely, we take $t_1 = T_0, t_2 = T_0 + \delta t, t_3 = T_0 + 2\delta t\cdots, t_M = T_1$ for time-average TE-QSCI, where $M = 1 + (T_1-T_0)/ \delta t$ is the total number of times.
In this case $\delta t$ is chosen equal to the Trotter step size, but in general it can be larger than it.
The shot simulation is employed and the total number of shots for measuring all times in the range is set to $8.85 \times 10^{5}$. The dimension of the subspace is taken as $R=850$.
The results in Table~\ref{tab:multipletimecomparison} illustrate that the performance of time-average TE-QSCI is comparable to single-time TE-QSCI at optimal time $t=1.4$ with the same number of shots and $R$ at the chosen time ranges. This observation is practically useful as we do not know the optimal time \textit{a priori}, and this implies that accurate results can be obtained by averaging over times in a finite range.

%%% ----- %%%

\subsection{Performance comparison using the UCCSD ansatz with classical CCSD amplitudes \label{subsec:comparison with CCSD}}

\begin{table*}[htb!]
\begin{tabular}{|l|l|l|l|l|l|l|l|l|l|}
\hline
Molecule & \ce{H6} & \ce{H8}  & \ce{H10} & \ce{N2}(8o,10e)  & \ce{N2}(10o,14e)  & \ce{NH3} & \ce{C2H2}  & Benzene (10e,10o) & \ce{CH4}  \\ \hline
$|\langle \psi_{GS}| \psi_{UCCSD} \rangle|^2 $ & 0.9996 &  0.9989 & 0.9979  & 0.9988 & 0.9988 & 0.9827 & 0.9996 & 0.9975 & 0.9999 \\ \hline
$E_\text{CCSD}-E_\text{exact}$ [mHa] & 0.389 &  1.073 & 1.980  & 4.517  & 4.518  & 0.210 &  2.186 & -28.433 & 0.225 \\ \hline
$R_{\text{UCCSD-QSCI}}$ (Trotter steps)  & 85 (1) & 1070 (10) & 11261 (20)  & -  &  -  & 77 (1)&  1284 (2) &  - & 174 (1) \\ \hline
$R_{\text{HF-TE-QSCI}}$ (Trotter steps)   & 87 (7) &  781 (7)  & 5830 (7)  & 128 (5) & 168 (5) & 100 (7) & 1448 (4)  & 99 (13)  & 222 (4) \\ \hline
$R_{\text{UCCSD-TE-QSCI}}$ (Trotter steps)  & 85 (2) & 696 (5) & 5072 (5)   & 118 (3) & 133 (3) & 81 (2)  & 977 (3) & 97 (12) & 171 (5) \\ \hline
\end{tabular}
\caption{
Comparison of the subspace dimension $R$ required to obtain $E_{\text{QSCI}}-E_{\text{exact}}<10^{-3}$ Hartree accuracy for the various methods.
The energy error of the classical CCSD calculation $E_\text{CCSD}-E_{\text{exact}}$ and the fidelity between the ground state and the UCCSD ansatz with the CCSD amplitudes $|\langle \psi_{GS}| \psi_{UCCSD} \rangle|^2$ are also displayed in the first two rows.
$R_\text{UCCSD-QSCI}, R_\text{HF-TE-QSCI}$, and $R_\text{UCCSD-TE-QSCI}$ represent the original QSCI for inputting the UCCSD ansatz with CCSD amplitudes, single-time TE-QSCI with the Hartree-Fock initial state, and single-time TE-QSCI taking the UCCSD ansatz with CCSD amplitudes as the initial state, respectively. The minus sign for some molecules in the UCCSD-QSCI column indicates that $10^-3$ accuracy cannot be obtained by sampling this state using a cutoff for the Born probabilities of $10^{-10}$. For HF-TE-QSCI, the Trotter step size is set to $\Delta t = 0.2$, while it is set to $\Delta t = 0.1$ for UCCSD-TE-QSCI. The total number of Trotter steps used in each method is denoted in the parenthesis in the last three lines of the table from which the optimal times can be inferred (note that this number includes one Trotter step for the UCCSD ansatze in UCCSD-TE-QSCI).}
\label{tab:UCCSDcomparison}
\end{table*}

So far we have considered the Hartree-Fock state as the initial state for TE-QSCI (which we will refer to as HF-TE-QSCI) for simplicity and as a proof of concept. TE-QSCI, however, is a general framework and we can consider more sophisticated initial states. 
Particularly, we consider the UCCSD ansatz state~\cite{Anand2022} with Trotterization whose parameters are determined by the classical CCSD calculation as an alternative initial state of TE-QSCI.
We also consider the result of QSCI just inputting this initial state (no time evolution) for performance comparison.

The UCCSD ansatz has been featured in applications of quantum computing to quantum chemistry and it can potentially express the approximate ground state by tuning its parameters.
The definition of the UCCSD ansatz written in fermion operators we adopt in this study is as follows (implemented in the package Quri Parts~\cite{quri-parts}:
\begin{align}
 \hat{U}_\mathrm{UCCSD}(\bmth)\ket{\mu_\mr{HF}} &= \mathrm{e}^{\hat{T}_s - \hat{T}_s^\dag} \mathrm{e}^{\hat{T}_{d1} - \hat{T}_{d1}^\dag} \mathrm{e}^{\hat{T}_{d2} - \hat{T}_{d2}^\dag} \ket{\mu_\mr{HF}}, \\
 \hat{T}_s &= \sum_{o,v} \theta_{ov}^{(s)} \left(  \hat{c}_{v\uparrow}^\dag \hat{c}_{o\uparrow} + \hat{c}_{v\downarrow}^\dag \hat{c}_{o\downarrow} \right),\\
 \hat{T}_{d1} &= \sum_{o, v} \theta^{(d1)}_{ov} \hat{c}_{v\uparrow}^\dag \hat{c}_{o\uparrow} \hat{c}_{v\downarrow}^\dag \hat{c}_{o\downarrow}, \\
 \hat{T}_{d2} &= \sum_{(o_1, v_1) \neq (o_2, v_2)} \sum_{\sigma, \tau = \uparrow, \downarrow} \theta_{o_1 o_2 v_1 v_2}^{(d2)} \hat{c}_{v_1\sigma}^\dagger \hat{c}_{o_1\sigma} \hat{c}_{v_2\tau}^\dagger \hat{c}_{o_2\tau},
\end{align}
where $\ket{\mu_\mr{HF}}$ is the Hartree-Fock state, $o,o_1,o_2$ ($v,v_1,v_2$) are the occupied (virtual) orbitals in the Hartree-Fock state, and $\bm{\theta} = \{\theta_{ov}^{(s)}, \theta_{o'v'}^{(d1)}, \theta_{o_1o_2v_1v_2}^{(d2)} | o,o',o_1,o_2 \in \text{occupied orbitals, } v,v',v_1,v_2 \in \text{virtual orbitals}\}$ are the circuit parameters.
When implementing this as a quantum circuit, we transform the fermion operators $\hat{c}, \hat{c}^\dag$ in the exponentials into qubit operators using the Jordan-Wigner transformation and approximate those exponentials by Trotterization.
The UCCSD ansatz is similar to the time-evolution operator $e^{-i\hat{H}t}$ used in TE-QSCI except that it includes only the excitations from the occupied to virtual orbitals and its parameters must be tuned by some optimization.
The optimization of the parameters is typically performed by expectation value estimation of the ansatz state, which requires a huge number of measurements in practice~\cite{Tilly2022}.
Nevertheless, it is suggested that the CCSD amplitudes obtained by the classical CCSD calculation are a good initial guess for the parameters~\cite{Hirsbrunner2024beyondmp}.
Therefore, it is natural to use the UCCSD ansatz with the CCSD amplitudes as an input state of QSCI because such a state could have a large overlap with the exact ground state.
We call QSCI with inputting the UCCSD ansatz with the CCSD amplitudes UCCSD-QSCI.
It should be noted that the large-scale hardware experiment of QSCI~\cite{robledomoreno2024} adopted a local version of the unitary cluster Jastrow ansatz~\cite{Matsuzawa2020, motta2023Bridging}, which approximates the UCCSD ansatz with the CCSD amplitudes by the singular value decomposition of the CCSD amplitudes.

While it is interesting to compare HF-TE-QSCI with UCCSD-QSCI to understand how well the naive form of TE-QSCI compares to sampling the classically approximated state, applying time-evolution to the UCCSD initial state (we will refer to this as UCCSD-TE-QSCI) can potentially lead to much better results. Indeed, as the fidelity between the UCCSD ansatz and ground state is higher than the fidelity between the Hartree-Fock and ground state, we would expect UCCSD-TE-QSCI to work better than HF-TE-QCSI based on the infinite time-average in Eq.~\eqref{eq:infinitetimeaverage}. 

In Table~\ref{tab:UCCSDcomparison}, we compare the subspace dimension $R$ required to obtain $E_{\text{QSCI}}-E_{\text{exact}}<10^{-3}$ Hartree accuracy for UCCSD-QSCI, single-time HF-TE-QSCI at the optimal time and single-time UCCSD-TE-QSCI at the optimal time. In addition to the molecules considered in the rest of the text, we investigate \ce{C2H2},\ce{CH4} and Benzene to obtain more data points for the comparison.
For these additional molecules, the atomic coordinates are taken from the CCCBDB database~\cite{NIST_CCCBDB} as the most stable structure at the level of Hartree-Fock/STO-3G and the Hamiltonians are prepared by using the Hartree-Fock orbital calculated by the STO-3G basis.
We take the active space approximation for benzene by taking 10 orbital and 10 electrons around HOMO and LUMO.
The numbers of qubits for \ce{C2H2},\ce{CH4} and Benzene are 24, 18, and 20, respectively.
Trotterization of 4, 5, 7 or 13 steps is used for single-time HF-TE-QSCI at the optimal times. To minimize the quantum resources required to prepare the initial state, we always Trotterize the UCCSD ansatz with a single Trotter step for UCCSD-TE-QSCI. The time evolution for UCCSD-TE-QSCI  uses a Trotterisation of 1, 2, 4 or 11 Trotter steps at the optimal times.  For UCCSD-QSCI, the number of Trotter steps for the UCCSD ansatz is determined by monitoring the convergence of the output QSCI energy when increasing the number of steps. This gives only a single step for most molecules, with the exception of \ce{H8} and \ce{H10} needing 10 and 20 steps, respectively. 

As seen in Table~\ref{tab:UCCSDcomparison}, the dimension $R$ required for 1 mHa accuracy is much larger for UCCSD-QSCI than that of HF-TE-QSCI for \ce{H8} and \ce{H10}, while it is slightly smaller for \ce{H6},\ce{NH3}, \ce{CH4} and \ce{C2H2}.
For \ce{N2} and Benzene, we found that it is impossible to obtain 1 mHa accuracy by sampling the UCCSD-QSCI state using a cutoff for the Born probabilities of $10^{-10}$.
We note that the Born probabilities for the smallest amplitudes required in (HF- and UCCSD-) TE-QSCI were around $10^{-5}$ in our simulations.

Our observations are summarized as follows.
We notice that the CCSD energy is already precise for \ce{H6}, \ce{NH3}, and \ce{CH4} ($ \ll 1$ mHa), where HF-TE-QSCI is slightly worse than UCCSD-QSCI, and that very compact subspaces achieve the desired accuracy regardless of the methods.
Inversely, these results may indicate that single-time HF-TE-QSCI can outperform UCCSD-QSCI when the problems are large and/or more complicated and the CCSD calculation is not so accurate (the error is $\gtrsim 1$mHa).
This expectation is, however, not the case for $\ce{C2H2}$, despite the CCSD calculation having an error of $\gtrsim 1$ mHa.
At least, the safe conclusion drawn from Table~\ref{tab:UCCSDcomparison} is that UCCSD-TE-QSCI always outperforms both HF-TE-QSCI and UCCSD-QSCI when the CCSD calculation has an error $\gtrsim 1$ mHa (and is essentially equivalent to UCCSD-QSCI when $ \ll 1$ mHa). 

We therefore see that the time evolution is a valid tool for preparing improved input states for QSCI, with even the simple HF-TE-QSCI outperforming UCCSD-QSCI in some cases. In particular, it can also provide compact subspaces for obtaining accurate energy in cases where UCCSD-QSCI cannot, such as \ce{N2} and Benzene.
Additionally, we observe that UCCSD-TE-QSCI always improves upon UCCSD-QSCI and HF-TE-QSCI or gives similar results.
Compared to HF-TE-QSCI, this is at the cost of additional quantum gates for preparing the initial state (the UCCSD ansatz) in addition to those for the time-evolution operator, but this essentially corresponds to adding a single time step and would generally be worth it. In Appendix~\ref{app:UCCSDtimeevolution}, we investigate the time dependence of UCCSD-TE-QSCI in more detail.

%%% ----- %%%
\subsection{System-size scaling of classical and quantum resources for TE-QSCI \label{subsec:scaling analysis}}
\begin{figure*}[htb!]
\includegraphics[width=1\linewidth]{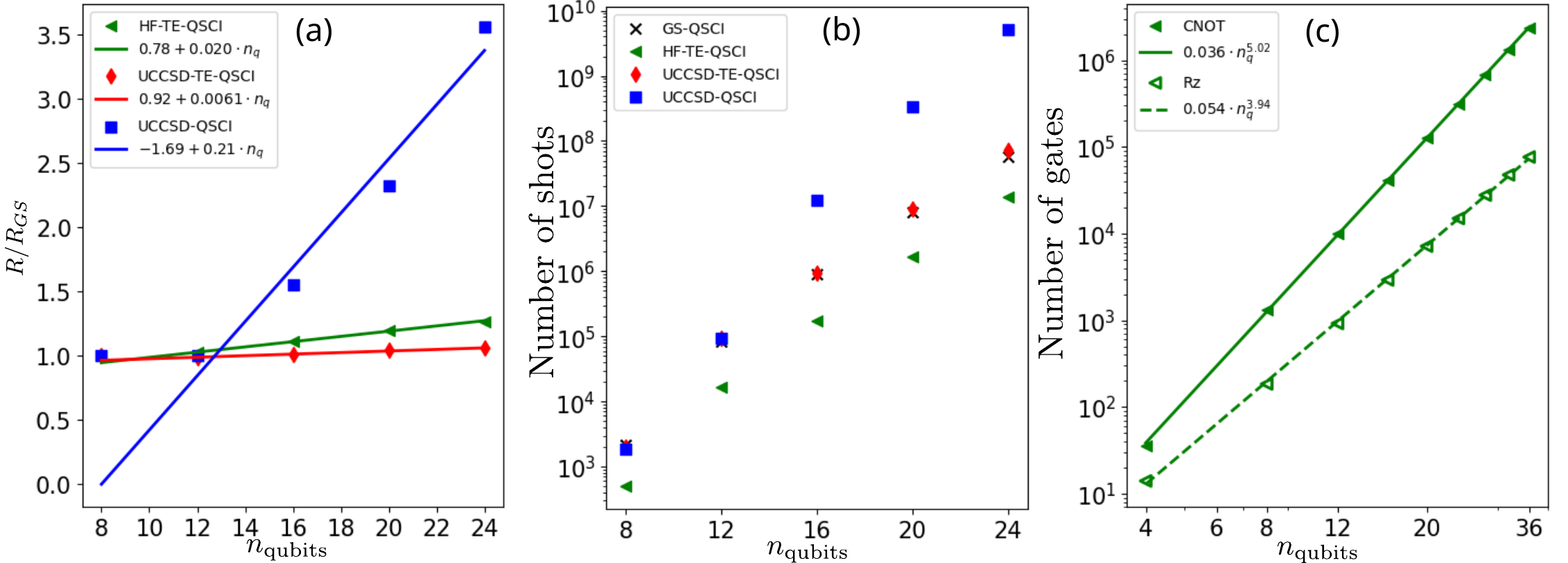}
\caption{System-size scaling of the classical and quantum resources of TE-QSCI.
(a): The ratio $R/R_{GS}$, where $R_{GS}$ is the subspace dimensions required for $10^{-3}$ Hartree accuracy for GS-QSCI, while $R$ are the subspace dimensions required for TE-QSCI with the Hartree-Fock initial state (HF-TE-QSCI) at $t=1.4$, TE-QSCI with the UCCSD initial state at $t=0.4$ (UCCSD-TE-QSCI) and UCCSD-QSCI. The lines represent linear fits.
(b): The estimate of the required number of shots to achieve $10^{-3}$ Hartree accuracy, corresponding to the results in (a).
(c): The number of gates required to implement a single Trotter step of the time-evolution operator $e^{-i\hath t}$. The lines show power-law fits.
}
\label{fig:Hydrogenscalingcomparison} 
\end{figure*}

Finally, we investigate the system-size scaling of TE-QSCI focusing on its classical and quantum resources.
This is important to tell whether our proposed method can tackle system sizes that cannot be exactly solved using classical methods.
Specifically, we consider hydrogen chain molecules systematically increasing the number of hydrogen atoms or qubits.
We construct the Hamiltonians [Eq.~\eqref{eq:Hamiltonian}] for \ce{H}$_n$ molecules where $n$ hydrogen atoms are aligned in a straight line with the atomic distance being 1 \AA{} by using the Hartree-Fock orbital with the STO-3G basis set.
The number of qubits for the Hamiltonian for the \ce{H}$_n$ molecule is $n_q=2n$.
We consider both HF-TE-QSCI, UCCSD-TE-QSCI (single Trotter-step implementation of UCCSD ansatz) and UCCSD-QSCI (exact implementation of UCCSD ansatz). The time-evolution operator is implemented exactly. 

As the classical resource requirements for TE-QSCI, we investigate how the dimension of the subspace $R$ required to obtain accurate results scales with system size.
Similarly to Sec.~\ref{subsec:performance of single TE QSCI}, we use the value of $R_{GS}$, which is defined such that GS-QSCI yields the sufficiently accurate ground-state energy whose error to the exact one is smaller than $10^{-3}$ Hartree, as the reference to discuss $R$ for the various input states. Note that the original proposal of QSCI~\cite{kanno2023quantum} studied $R_{GS}$ for hydrogen chains in detail. We once again stress that GS-QSCI is not an actual protocol that can be implemented on a quantum device and we use it as a benchmark for the comparison.
This is because GS-QSCI leads to a compact set of truncated bases (configurations) although they are not guaranteed to be optimal. If a given input state performs significantly worse than GS-QSCI, it can be considered not useful.

In Fig.~\ref{fig:Hydrogenscalingcomparison}(a), we show the ratio between the $R$ required for $10^{-3}$ Hartree accuracy for the various QSCI input states and GS-QSCI ($R_{GS}$) as a function of the number of qubits.
For a small system there is almost no added cost, but from $n_q\geq 12$ there is a linear increase in the ratio with proportionality $0.21 \cdot n_q$ for UCCSD-QSCI, $0.02 \cdot n_q$ for HF-TE-QSCI and $0.0061 \cdot n_q$ for UCCSD-TE-QSCI. The extrapolation to $n_q=36$ (50) leads to $R/R_{GS} = 5.87$ (8.81) for UCCSD-QSCI, $R/R_{GS} = 1.5$ (1.78) for HF-TE-QSCI and $R/R_{GS} = 1.14$ (1.23) for UCCSD-TE-QSCI. UCCSD-QSCI scales significantly worse than TE-QSCI with the linear coefficient being an order of magnitude larger than even HF-TE-QSCI and requiring 5-10 times more classical resources for 50 qubits compared to UCCSD-TE-QSCI and HF-TE-QSCI. Using this as the input state for QSCI is therefore highly nonoptimal and doesn't provide a compact Hilbert space.
On the other hand, the additional classical computational cost of TE-QSCI compared with GS-QSCI, which uses one of the optimal input states for QSCI, is not so large even for 50 qubits. This tendency is particularly evident for UCCSD-TE-QSCI, which improves the scaling significantly compared to HF-TE-QSCI, although it requires a classical precalculation (CCSD) and the implementation of the UCCSD ansatz before the time evolution.

Next, we consider the sampling efficiency of TE-QSCI. This is one of the basic quantum resource requirements as it corresponds to the number of circuit repetitions required. The number of required shots for (the fictitious) GS-QSCI was discussed in the original proposal of QSCI~\cite{kanno2023quantum} and generally it is comparable to or smaller than the number of shots to estimate the Hamiltonian expectation value in the standard protocol \textit{once}.
This means that GS-QSCI is much more shot-efficient compared with other algorithms using the Hamiltonian expectation value, such as VQE or the quantum Krylov method.

We estimate the number of required shots for QSCI with various initial states by taking the inverse of the smallest Born probability among the $R$ configurations picked up by QSCI.
This corresponds to the smallest probability of the configurations that must be sampled by QSCI, so the inverse of it gives a rough estimate of the required number of shots to find $R$ configurations by QSCI. The results are shown in Fig.~\ref{fig:Hydrogenscalingcomparison}(b).
The number of samples required for UCCSD-TE-QSCI is almost the same as that for GS-QSCI, while HF-TE-QSCI requires less samples compared to these two.
Because UCCSD-QSCI has to pick out $2-3.5$ times more basis states, it requires the resolution of much smaller probabilities and therefore at least an order of magnitude more measurements. Interestingly HF-TE-QSCI needs a smaller number of quantum measurements than UCCSD-TE-QSCI, despite also needing to find a slightly larger amount of basis states (subspace dimension $R$).

Another crucial aspect of the quantum resource requirements of TE-QSCI is the number of elementary gates required to run the TEO circuit $e^{-i\hat{H}t}$ once. The elementary gates considered in our estimation are the single-qubit $z$-rotation gate $R_z(\phi)$ with arbitrary angle $\phi$, the CNOT gate, and single-qubit Clifford gates.
The number of CNOT gates is typically more important than that of single-qubit gates in the NISQ era because the fidelity of the former is much worse than the latter in currently available NISQ hardware.
On the other hand, the number of rotation gates is crucial and that of CNOT becomes marginal when we have (early-)FTQC era in mind since the Clifford gates, including CNOT, are relatively easy to implement in FTQC and the non-Clifford gates such as $R_z(\phi)$ are the most cost-consuming.
To obtain a rough estimate of the number of these two gates for the TEO circuit in actual implementations, we consider the approximated TEO circuit which is Trotterized with a single step. 
The circuit consists of a sequence of the rotation gates by Pauli operators acting on $n_q$ qubits, and we decompose those rotation gates into CNOT gates and $Rz(\phi)$ gates by the standard way~\cite{Anand2022}.
Namely, when the weight of the Pauli operator, or the number of qubits on which the Pauli operator acts non-trivially, is $p$, the rotation gate is decomposed into $2(p-1)$ CNOT gates and one $Rz(\phi)$ gate with other single-qubit Clifford gates.
Note that this estimation assumes all-to-all connectivity among qubits.

In Fig.~\ref{fig:Hydrogenscalingcomparison}(c), we show the numerical counts of the gates for implementing the TEO circuit with the single-step Trotterization (independent of the time of the TEO circuit).
The counts obey a power law with the number of qubits, and the numerical fit gives the exponents as $a_{\text{CNOT}}=5.02$ and $a_{R_z}=3.94$.
This is consistent with the generic expectation that the number of CNOT ($R_z(\phi)$) scales with $n_q^5$ ($n_q^4$) where $n_q$ is the number of qubits~\cite{Anand2022}, explained by the fact that the TEO circuit (similar to the UCCSD anzats) has $O(n_q^4)$ gates and each gate is decomposed into $O(n_q)$ CNOTs and one $R_z(\phi)$ gate on average.
The number of CNOT gates for a single Trotter step is on the order of $10^6$ for the 36-qubit system where naive classical exact diagonalization becomes difficult.
This count is rather large considering that the current quantum hardware often exhibits an error rate of $\sim 10^{-3}$ for two-qubit gates, so performing TE-QSCI on NISQ devices for classically intractable systems would need more improvement of the algorithm, quantum circuit, etc. On the other hand, the number of $R_z(\phi)$ gates is on the order of $10^4$ for the 36-qubit system. This value is close to the maximal number of non-Clifford operations in the recently proposed early-FTQC architecture called STAR~\cite{Akahoshi2024}, so TE-QSCI would be feasible on such devices. Note that we do not simplify the circuit, so there is much room for improving the estimate gate counts.

\section{Summary, discussion and outlook}
\label{sec:discsussion}
In this study, we have shown that applying the time-evolution operator to suitable initial states can generate promising candidate states for QSCI to calculate accurate ground-state energies of quantum chemistry Hamiltonians. 
We have investigated how the accuracy of this protocol, Time-Evolved QSCI (TE-QSCI), depends on the time, Trotter step size and the number of basis states $R$ in the selected subspace, using the Hartree-Fock state as the initial state (HF-TE-QSCI) and taking small molecules as examples.
All investigated cases exhibit similar optimal time ranges where the highest accuracy for single-time TE-QSCI is obtained when $R$ is kept fixed.
Searching for the optimal time may require performing single-time TE-QSCI for multiple times and a large number of measurements, but time-average TE-QSCI is possibly useful to obtain the accurate energy without knowing the exact optimal time.
In general, we also find that there is a trade-off between quantum and classical resources: Increasing the Trotter step size thereby reducing the number of gates in quantum circuits leads to more unimportant electron configurations in the selected subspace, but this can be compensated by increasing $R$ at the expense of a higher classical computational cost for exact diagonalization.

In addition to HF-TE-QSCI, we also compared with two other input states of QSCI based on utilizing the UCCSD ansatz with the classically calculated CCSD amplitudes. One was using this state directly as the input of QSCI (UCCSD-QSCI) and the second was using its time evolution as the input to QSCI (UCCSD-TE-QSCI).
The numerical results indicate that TE-QSCI with the Hartree-Fock initial state and the UCCSD ansatz can potentially give more accurate results than UCCSD-QSCI, which in some cases (\ce{N2} and Benzene) cannot find compact subspaces to calculate the sufficiently accurate energy. Additionally, we found that UCCSD-TE-QSCI performs better than or similar to UCCSD-QSCI for all molecules checked.

All of these input methods have an intrinsic overhead in terms of the number of states $R$ required for a given accuracy in comparison with the number of states $R_{GS}$ required when inputting the exact ground state to QSCI (GS-QSCI). Note that GS-QSCI is not an actual protocol and cannot be implemented on a quantum device, it is soley used as a reference. 
Systematic analysis of the system-size scaling using hydrogen chains represented by $n_q$ qubits revealed that the scaling of the overhead
$R/R_{GS}$ was numerically found to be $0.21 \cdot n_q$ for UCCSD-QSCI, $0.02 \cdot n_q$ for HF-TE-QSCI and $0.0061 \cdot n_q$ for UCCSD-TE-QSCI. The classical resources required for UCCSD-QSCI scale significantly worse than the two TE-QSCI methods and also required at least an order of magnitude more measurements, underscoring the practical utility of using time-evolution circuits compared to sampling based purely on classical approximations of the ground state. 
Indeed TE-QSCI is not prohibitive for practical application to large systems giving a value of, e.g., $R/R_{GS}=1.78$ for HF-TE-QSCI and $1.23$ for UCCSD-TE-QSCI at 50 qubits. The scaling analysis on the number of quantum gates to implement the quantum circuit for TE-QSCI implies that the gate counts for classically intractable problems is within the capability of early-FTQC devices such as that proposed in Ref.~\cite{Akahoshi2024} though it seems difficult to run on current NISQ devices in the naive form. 

Overall, TE-QSCI can serve as a practical way to implement QSCI for calculating an accurate ground state without optimizing the quantum circuit by quantum-classical feedback. In particular, both HF- and UCCSD-TE-QSCI perform much better than UCCSD-QSCI for the hydrogen chain, while UCCSD-TE-QSCI performs better than or similar to UCCSD-QSCI for all molecules checked.

Note that UCCSD-QSCI considered in this paper would be more accurate in the noiseless limit than performing QSCI with inputting the local version of the unitary cluster Jastrow ansatz as done in \cite{robledomoreno2024}, because the latter is a rough approximation of the former with a shallower quantum circuit suited for NISQ devices.

There are several interesting future directions of this study.
First, it is vital to reduce the number of gates in Fig.~\ref{fig:Hydrogenscalingcomparison}(c) if we want to run TE-QSCI on actual NISQ hardware.
For example, by relaxing the implementation of the time-evolution operator $e^{-i\hat{H}t}$ into a more approximated one than the Trotterization, like using the stochastic application of the gates~\cite{campbel2019} or dropping the gates depending on the hardware requirement~\cite{motta2023Bridging, robledomoreno2024}.
We stress that the estimation in Fig.~\ref{fig:Hydrogenscalingcomparison}(c) does not take any simplification of quantum circuits into account.
Second, searching for an initial state of TE-QSCI other than the Hartree-Fock state or the UCCSD ansatz is also intriguing because an initial state with a larger ground-state fidelity can improve the accuracy of QSCI for a given subspace dimension, as shown in Sec.~\ref{subsec:comparison with CCSD} and Appendix~\ref{app:fidelitydependence}.

While the use of the UCCSD initial state for TE-QSCI improves the classical resource requirement of the algorithm (Fig.~\ref{fig:Hydrogenscalingcomparison}(a)), it requires the implementation of the UCCSD ansatz as a quantum circuit.
In Secs.~\ref{subsec:comparison with CCSD} and \ref{subsec:scaling analysis}, we consider the UCCSD ansatz with single-step Trotterization, so the additional gates required for the UCCSD initial state compared to the Hartree-Fock initial state approximately corresponds to adding just a single time step to the time-evolution operator circuit,  because the UCCSD ansatz has a similar form to the time-evolution operator (i.e., an exponential of single- and double-electron excitations).
Additionally, the optimal time of TE-QSCI becomes smaller, which decreases the number of Trotter steps required for the time-evolution operator (see Appendix~\ref{app:UCCSDtimeevolution}). 
A more detailed understanding of the classical and quantum resources and overall generalizability of UCCSD-TE-QSCI is therefore of interest.
We also point out that another potential way to prepare an initial state of TE-QSCI is to run VQE or ADAPT-QSCI for a limited number of optimization steps, using the obtained circuit as the initial state.
Third, as a related research direction, the application of QSCI to calculate the dynamics of quantum systems is important. By picking out important configurations at each time during the time evolution by measuring the time-evolved state, we can construct a time-dependent subspace to be used for classical time-evolution methods.
\section*{Note added}
In the final stage of preparing this manuscript, Sugisaki \textit{et al.} independently posted a manuscript to arXiv~\cite{sugisaki2024}, which utilizes the same idea as this study and named their method the Hamiltonian-simulation based QSCI (\textit{HSB-QSCI}).
They considered the time-evolved states as inputs of QSCI and the dimension of the subspace was not fixed, whereas our proposal allows a more general framework and investigates the effect of limiting the maximum size of the subspace dimension.
Their manuscript presents an interesting study focusing on the practical application to oligoacenes and includes an actual hardware experiment on IBM quantum devices. Our study has a different focus, namely investigating the optimality of the configurations chosen by TE-QSCI (or HSB-QSCI) in order to address how well the classical resource requirements of the method scales to large systems. We believe the two manuscripts are complimentary.

\section*{Acknowledgment}
The authors thank Keita Kanno for initiation of this research project and helpful discussion.

%%% --------------- %%%
\bibliography{ref} 

\begin{thebibliography}{66}%
\makeatletter
\providecommand \@ifxundefined [1]{%
 \@ifx{#1\undefined}
}%
\providecommand \@ifnum [1]{%
 \ifnum #1\expandafter \@firstoftwo
 \else \expandafter \@secondoftwo
 \fi
}%
\providecommand \@ifx [1]{%
 \ifx #1\expandafter \@firstoftwo
 \else \expandafter \@secondoftwo
 \fi
}%
\providecommand \natexlab [1]{#1}%
\providecommand \enquote  [1]{``#1''}%
\providecommand \bibnamefont  [1]{#1}%
\providecommand \bibfnamefont [1]{#1}%
\providecommand \citenamefont [1]{#1}%
\providecommand \href@noop [0]{\@secondoftwo}%
\providecommand \href [0]{\begingroup \@sanitize@url \@href}%
\providecommand \@href[1]{\@@startlink{#1}\@@href}%
\providecommand \@@href[1]{\endgroup#1\@@endlink}%
\providecommand \@sanitize@url [0]{\catcode `\\12\catcode `\$12\catcode
  `\&12\catcode `\#12\catcode `\^12\catcode `\_12\catcode `\%12\relax}%
\providecommand \@@startlink[1]{}%
\providecommand \@@endlink[0]{}%
\providecommand \url  [0]{\begingroup\@sanitize@url \@url }%
\providecommand \@url [1]{\endgroup\@href {#1}{\urlprefix }}%
\providecommand \urlprefix  [0]{URL }%
\providecommand \Eprint [0]{\href }%
\providecommand \doibase [0]{https://doi.org/}%
\providecommand \selectlanguage [0]{\@gobble}%
\providecommand \bibinfo  [0]{\@secondoftwo}%
\providecommand \bibfield  [0]{\@secondoftwo}%
\providecommand \translation [1]{[#1]}%
\providecommand \BibitemOpen [0]{}%
\providecommand \bibitemStop [0]{}%
\providecommand \bibitemNoStop [0]{.\EOS\space}%
\providecommand \EOS [0]{\spacefactor3000\relax}%
\providecommand \BibitemShut  [1]{\csname bibitem#1\endcsname}%
\let\auto@bib@innerbib\@empty
%</preamble>
\bibitem [{\citenamefont {Feynman}(1982)}]{Feynman1982}%
  \BibitemOpen
  \bibfield  {author} {\bibinfo {author} {\bibfnamefont {R.~P.}\ \bibnamefont
  {Feynman}},\ }\bibfield  {title} {\bibinfo {title} {Simulating physics with
  computers},\ }\href {https://doi.org/10.1007/BF02650179} {\bibfield
  {journal} {\bibinfo  {journal} {International Journal of Theoretical
  Physics}\ }\textbf {\bibinfo {volume} {21}},\ \bibinfo {pages} {467}
  (\bibinfo {year} {1982})}\BibitemShut {NoStop}%
\bibitem [{\citenamefont {Cao}\ \emph {et~al.}(2019)\citenamefont {Cao},
  \citenamefont {Romero}, \citenamefont {Olson}, \citenamefont {Degroote},
  \citenamefont {Johnson}, \citenamefont {Kieferov{\'a}}, \citenamefont
  {Kivlichan}, \citenamefont {Menke}, \citenamefont {Peropadre}, \citenamefont
  {Sawaya}, \citenamefont {Sim}, \citenamefont {Veis},\ and\ \citenamefont
  {Aspuru-Guzik}}]{Cao2019}%
  \BibitemOpen
  \bibfield  {author} {\bibinfo {author} {\bibfnamefont {Y.}~\bibnamefont
  {Cao}}, \bibinfo {author} {\bibfnamefont {J.}~\bibnamefont {Romero}},
  \bibinfo {author} {\bibfnamefont {J.~P.}\ \bibnamefont {Olson}}, \bibinfo
  {author} {\bibfnamefont {M.}~\bibnamefont {Degroote}}, \bibinfo {author}
  {\bibfnamefont {P.~D.}\ \bibnamefont {Johnson}}, \bibinfo {author}
  {\bibfnamefont {M.}~\bibnamefont {Kieferov{\'a}}}, \bibinfo {author}
  {\bibfnamefont {I.~D.}\ \bibnamefont {Kivlichan}}, \bibinfo {author}
  {\bibfnamefont {T.}~\bibnamefont {Menke}}, \bibinfo {author} {\bibfnamefont
  {B.}~\bibnamefont {Peropadre}}, \bibinfo {author} {\bibfnamefont {N.~P.~D.}\
  \bibnamefont {Sawaya}}, \bibinfo {author} {\bibfnamefont {S.}~\bibnamefont
  {Sim}}, \bibinfo {author} {\bibfnamefont {L.}~\bibnamefont {Veis}},\ and\
  \bibinfo {author} {\bibfnamefont {A.}~\bibnamefont {Aspuru-Guzik}},\
  }\bibfield  {title} {\bibinfo {title} {Quantum chemistry in the age of
  quantum computing},\ }\href {https://doi.org/10.1021/acs.chemrev.8b00803}
  {\bibfield  {journal} {\bibinfo  {journal} {Chemical Reviews}\ }\textbf
  {\bibinfo {volume} {119}},\ \bibinfo {pages} {10856} (\bibinfo {year}
  {2019})}\BibitemShut {NoStop}%
\bibitem [{\citenamefont {McArdle}\ \emph {et~al.}(2020)\citenamefont
  {McArdle}, \citenamefont {Endo}, \citenamefont {Aspuru-Guzik}, \citenamefont
  {Benjamin},\ and\ \citenamefont {Yuan}}]{McArdle2020}%
  \BibitemOpen
  \bibfield  {author} {\bibinfo {author} {\bibfnamefont {S.}~\bibnamefont
  {McArdle}}, \bibinfo {author} {\bibfnamefont {S.}~\bibnamefont {Endo}},
  \bibinfo {author} {\bibfnamefont {A.}~\bibnamefont {Aspuru-Guzik}}, \bibinfo
  {author} {\bibfnamefont {S.~C.}\ \bibnamefont {Benjamin}},\ and\ \bibinfo
  {author} {\bibfnamefont {X.}~\bibnamefont {Yuan}},\ }\bibfield  {title}
  {\bibinfo {title} {Quantum computational chemistry},\ }\href
  {https://doi.org/10.1103/RevModPhys.92.015003} {\bibfield  {journal}
  {\bibinfo  {journal} {Rev. Mod. Phys.}\ }\textbf {\bibinfo {volume} {92}},\
  \bibinfo {pages} {015003} (\bibinfo {year} {2020})}\BibitemShut {NoStop}%
\bibitem [{\citenamefont {Kitaev}(1995)}]{kitaev1995quantum}%
  \BibitemOpen
  \bibfield  {author} {\bibinfo {author} {\bibfnamefont {A.~Y.}\ \bibnamefont
  {Kitaev}},\ }\bibfield  {title} {\bibinfo {title} {Quantum measurements and
  the abelian stabilizer problem},\ }\href@noop {} {\bibfield  {journal}
  {\bibinfo  {journal} {arXiv preprint quant-ph/9511026}\ } (\bibinfo {year}
  {1995})}\BibitemShut {NoStop}%
\bibitem [{\citenamefont {Cleve}\ \emph {et~al.}(1998)\citenamefont {Cleve},
  \citenamefont {Ekert}, \citenamefont {Macchiavello},\ and\ \citenamefont
  {Mosca}}]{cleve1998}%
  \BibitemOpen
  \bibfield  {author} {\bibinfo {author} {\bibfnamefont {R.}~\bibnamefont
  {Cleve}}, \bibinfo {author} {\bibfnamefont {A.}~\bibnamefont {Ekert}},
  \bibinfo {author} {\bibfnamefont {C.}~\bibnamefont {Macchiavello}},\ and\
  \bibinfo {author} {\bibfnamefont {M.}~\bibnamefont {Mosca}},\ }\bibfield
  {title} {\bibinfo {title} {Quantum algorithms revisited},\ }\href
  {https://doi.org/10.1098/rspa.1998.0164} {\bibfield  {journal} {\bibinfo
  {journal} {Proceedings of the Royal Society of London. Series A:
  Mathematical, Physical and Engineering Sciences}\ }\textbf {\bibinfo {volume}
  {454}},\ \bibinfo {pages} {339} (\bibinfo {year} {1998})}\BibitemShut
  {NoStop}%
\bibitem [{\citenamefont {Peruzzo}\ \emph {et~al.}(2014)\citenamefont
  {Peruzzo}, \citenamefont {McClean}, \citenamefont {Shadbolt}, \citenamefont
  {Yung}, \citenamefont {Zhou}, \citenamefont {Love}, \citenamefont
  {Aspuru-Guzik},\ and\ \citenamefont {O'Brien}}]{Peruzzo2014}%
  \BibitemOpen
  \bibfield  {author} {\bibinfo {author} {\bibfnamefont {A.}~\bibnamefont
  {Peruzzo}}, \bibinfo {author} {\bibfnamefont {J.}~\bibnamefont {McClean}},
  \bibinfo {author} {\bibfnamefont {P.}~\bibnamefont {Shadbolt}}, \bibinfo
  {author} {\bibfnamefont {M.-H.}\ \bibnamefont {Yung}}, \bibinfo {author}
  {\bibfnamefont {X.-Q.}\ \bibnamefont {Zhou}}, \bibinfo {author}
  {\bibfnamefont {P.~J.}\ \bibnamefont {Love}}, \bibinfo {author}
  {\bibfnamefont {A.}~\bibnamefont {Aspuru-Guzik}},\ and\ \bibinfo {author}
  {\bibfnamefont {J.~L.}\ \bibnamefont {O'Brien}},\ }\bibfield  {title}
  {\bibinfo {title} {A variational eigenvalue solver on a photonic quantum
  processor},\ }\href {https://doi.org/10.1038/ncomms5213} {\bibfield
  {journal} {\bibinfo  {journal} {Nature Communications}\ }\textbf {\bibinfo
  {volume} {5}},\ \bibinfo {pages} {4213} (\bibinfo {year} {2014})}\BibitemShut
  {NoStop}%
\bibitem [{\citenamefont {Tilly}\ \emph {et~al.}(2022)\citenamefont {Tilly},
  \citenamefont {Chen}, \citenamefont {Cao}, \citenamefont {Picozzi},
  \citenamefont {Setia}, \citenamefont {Li}, \citenamefont {Grant},
  \citenamefont {Wossnig}, \citenamefont {Rungger}, \citenamefont {Booth},\
  and\ \citenamefont {Tennyson}}]{Tilly2022}%
  \BibitemOpen
  \bibfield  {author} {\bibinfo {author} {\bibfnamefont {J.}~\bibnamefont
  {Tilly}}, \bibinfo {author} {\bibfnamefont {H.}~\bibnamefont {Chen}},
  \bibinfo {author} {\bibfnamefont {S.}~\bibnamefont {Cao}}, \bibinfo {author}
  {\bibfnamefont {D.}~\bibnamefont {Picozzi}}, \bibinfo {author} {\bibfnamefont
  {K.}~\bibnamefont {Setia}}, \bibinfo {author} {\bibfnamefont
  {Y.}~\bibnamefont {Li}}, \bibinfo {author} {\bibfnamefont {E.}~\bibnamefont
  {Grant}}, \bibinfo {author} {\bibfnamefont {L.}~\bibnamefont {Wossnig}},
  \bibinfo {author} {\bibfnamefont {I.}~\bibnamefont {Rungger}}, \bibinfo
  {author} {\bibfnamefont {G.~H.}\ \bibnamefont {Booth}},\ and\ \bibinfo
  {author} {\bibfnamefont {J.}~\bibnamefont {Tennyson}},\ }\bibfield  {title}
  {\bibinfo {title} {The variational quantum eigensolver: A review of methods
  and best practices},\ }\href
  {https://doi.org/https://doi.org/10.1016/j.physrep.2022.08.003} {\bibfield
  {journal} {\bibinfo  {journal} {Physics Reports}\ }\textbf {\bibinfo {volume}
  {986}},\ \bibinfo {pages} {1} (\bibinfo {year} {2022})}\BibitemShut {NoStop}%
\bibitem [{\citenamefont {McClean}\ \emph {et~al.}(2018)\citenamefont
  {McClean}, \citenamefont {Boixo}, \citenamefont {Smelyanskiy}, \citenamefont
  {Babbush},\ and\ \citenamefont {Neven}}]{McClean2018Barren}%
  \BibitemOpen
  \bibfield  {author} {\bibinfo {author} {\bibfnamefont {J.~R.}\ \bibnamefont
  {McClean}}, \bibinfo {author} {\bibfnamefont {S.}~\bibnamefont {Boixo}},
  \bibinfo {author} {\bibfnamefont {V.~N.}\ \bibnamefont {Smelyanskiy}},
  \bibinfo {author} {\bibfnamefont {R.}~\bibnamefont {Babbush}},\ and\ \bibinfo
  {author} {\bibfnamefont {H.}~\bibnamefont {Neven}},\ }\bibfield  {title}
  {\bibinfo {title} {Barren plateaus in quantum neural network training
  landscapes},\ }\href {https://doi.org/10.1038/s41467-018-07090-4} {\bibfield
  {journal} {\bibinfo  {journal} {Nature Communications}\ }\textbf {\bibinfo
  {volume} {9}},\ \bibinfo {pages} {4812} (\bibinfo {year} {2018})}\BibitemShut
  {NoStop}%
\bibitem [{\citenamefont {Gonthier}\ \emph {et~al.}(2022)\citenamefont
  {Gonthier}, \citenamefont {Radin}, \citenamefont {Buda}, \citenamefont
  {Doskocil}, \citenamefont {Abuan},\ and\ \citenamefont
  {Romero}}]{Gonthier2022}%
  \BibitemOpen
  \bibfield  {author} {\bibinfo {author} {\bibfnamefont {J.~F.}\ \bibnamefont
  {Gonthier}}, \bibinfo {author} {\bibfnamefont {M.~D.}\ \bibnamefont {Radin}},
  \bibinfo {author} {\bibfnamefont {C.}~\bibnamefont {Buda}}, \bibinfo {author}
  {\bibfnamefont {E.~J.}\ \bibnamefont {Doskocil}}, \bibinfo {author}
  {\bibfnamefont {C.~M.}\ \bibnamefont {Abuan}},\ and\ \bibinfo {author}
  {\bibfnamefont {J.}~\bibnamefont {Romero}},\ }\bibfield  {title} {\bibinfo
  {title} {Measurements as a roadblock to near-term practical quantum advantage
  in chemistry: Resource analysis},\ }\href
  {https://doi.org/10.1103/PhysRevResearch.4.033154} {\bibfield  {journal}
  {\bibinfo  {journal} {Phys. Rev. Res.}\ }\textbf {\bibinfo {volume} {4}},\
  \bibinfo {pages} {033154} (\bibinfo {year} {2022})}\BibitemShut {NoStop}%
\bibitem [{\citenamefont {Layden}\ \emph {et~al.}(2023)\citenamefont {Layden},
  \citenamefont {Mazzola}, \citenamefont {Mishmash}, \citenamefont {Motta},
  \citenamefont {Wocjan}, \citenamefont {Kim},\ and\ \citenamefont
  {Sheldon}}]{Layden2023}%
  \BibitemOpen
  \bibfield  {author} {\bibinfo {author} {\bibfnamefont {D.}~\bibnamefont
  {Layden}}, \bibinfo {author} {\bibfnamefont {G.}~\bibnamefont {Mazzola}},
  \bibinfo {author} {\bibfnamefont {R.~V.}\ \bibnamefont {Mishmash}}, \bibinfo
  {author} {\bibfnamefont {M.}~\bibnamefont {Motta}}, \bibinfo {author}
  {\bibfnamefont {P.}~\bibnamefont {Wocjan}}, \bibinfo {author} {\bibfnamefont
  {J.-S.}\ \bibnamefont {Kim}},\ and\ \bibinfo {author} {\bibfnamefont
  {S.}~\bibnamefont {Sheldon}},\ }\bibfield  {title} {\bibinfo {title}
  {Quantum-enhanced markov chain monte carlo},\ }\href
  {https://doi.org/10.1038/s41586-023-06095-4} {\bibfield  {journal} {\bibinfo
  {journal} {Nature}\ }\textbf {\bibinfo {volume} {619}},\ \bibinfo {pages}
  {282} (\bibinfo {year} {2023})}\BibitemShut {NoStop}%
\bibitem [{\citenamefont {Huggins}\ \emph {et~al.}(2022)\citenamefont
  {Huggins}, \citenamefont {O'Gorman}, \citenamefont {Rubin}, \citenamefont
  {Reichman}, \citenamefont {Babbush},\ and\ \citenamefont
  {Lee}}]{Huggins2022}%
  \BibitemOpen
  \bibfield  {author} {\bibinfo {author} {\bibfnamefont {W.~J.}\ \bibnamefont
  {Huggins}}, \bibinfo {author} {\bibfnamefont {B.~A.}\ \bibnamefont
  {O'Gorman}}, \bibinfo {author} {\bibfnamefont {N.~C.}\ \bibnamefont {Rubin}},
  \bibinfo {author} {\bibfnamefont {D.~R.}\ \bibnamefont {Reichman}}, \bibinfo
  {author} {\bibfnamefont {R.}~\bibnamefont {Babbush}},\ and\ \bibinfo {author}
  {\bibfnamefont {J.}~\bibnamefont {Lee}},\ }\bibfield  {title} {\bibinfo
  {title} {Unbiasing fermionic quantum monte carlo with a quantum computer},\
  }\href {https://doi.org/10.1038/s41586-021-04351-z} {\bibfield  {journal}
  {\bibinfo  {journal} {Nature}\ }\textbf {\bibinfo {volume} {603}},\ \bibinfo
  {pages} {416} (\bibinfo {year} {2022})}\BibitemShut {NoStop}%
\bibitem [{\citenamefont {Parrish}\ and\ \citenamefont
  {McMahon}(2019)}]{parrish2019quantum}%
  \BibitemOpen
  \bibfield  {author} {\bibinfo {author} {\bibfnamefont {R.~M.}\ \bibnamefont
  {Parrish}}\ and\ \bibinfo {author} {\bibfnamefont {P.~L.}\ \bibnamefont
  {McMahon}},\ }\bibfield  {title} {\bibinfo {title} {Quantum filter
  diagonalization: Quantum eigendecomposition without full quantum phase
  estimation},\ }\href@noop {} {\bibfield  {journal} {\bibinfo  {journal}
  {arXiv preprint arXiv:1909.08925}\ } (\bibinfo {year} {2019})}\BibitemShut
  {NoStop}%
\bibitem [{\citenamefont {Stair}\ \emph {et~al.}(2020)\citenamefont {Stair},
  \citenamefont {Huang},\ and\ \citenamefont {Evangelista}}]{Stair2020}%
  \BibitemOpen
  \bibfield  {author} {\bibinfo {author} {\bibfnamefont {N.~H.}\ \bibnamefont
  {Stair}}, \bibinfo {author} {\bibfnamefont {R.}~\bibnamefont {Huang}},\ and\
  \bibinfo {author} {\bibfnamefont {F.~A.}\ \bibnamefont {Evangelista}},\
  }\bibfield  {title} {\bibinfo {title} {A multireference quantum krylov
  algorithm for strongly correlated electrons},\ }\href
  {https://doi.org/10.1021/acs.jctc.9b01125} {\bibfield  {journal} {\bibinfo
  {journal} {Journal of Chemical Theory and Computation}\ }\textbf {\bibinfo
  {volume} {16}},\ \bibinfo {pages} {2236} (\bibinfo {year}
  {2020})}\BibitemShut {NoStop}%
\bibitem [{\citenamefont {Kanno}\ \emph {et~al.}(2023)\citenamefont {Kanno},
  \citenamefont {Kohda}, \citenamefont {Imai}, \citenamefont {Koh},
  \citenamefont {Mitarai}, \citenamefont {Mizukami},\ and\ \citenamefont
  {Nakagawa}}]{kanno2023quantum}%
  \BibitemOpen
  \bibfield  {author} {\bibinfo {author} {\bibfnamefont {K.}~\bibnamefont
  {Kanno}}, \bibinfo {author} {\bibfnamefont {M.}~\bibnamefont {Kohda}},
  \bibinfo {author} {\bibfnamefont {R.}~\bibnamefont {Imai}}, \bibinfo {author}
  {\bibfnamefont {S.}~\bibnamefont {Koh}}, \bibinfo {author} {\bibfnamefont
  {K.}~\bibnamefont {Mitarai}}, \bibinfo {author} {\bibfnamefont
  {W.}~\bibnamefont {Mizukami}},\ and\ \bibinfo {author} {\bibfnamefont
  {Y.~O.}\ \bibnamefont {Nakagawa}},\ }\bibfield  {title} {\bibinfo {title}
  {{Quantum-Selected Configuration Interaction: classical diagonalization of
  Hamiltonians in subspaces selected by quantum computers}},\ }\href@noop {}
  {\bibfield  {journal} {\bibinfo  {journal} {arXiv preprint arXiv:2302.11320}\
  } (\bibinfo {year} {2023})}\BibitemShut {NoStop}%
\bibitem [{\citenamefont {Arute}\ \emph {et~al.}(2019)\citenamefont {Arute},
  \citenamefont {Arya}, \citenamefont {Babbush}, \citenamefont {Bacon},
  \citenamefont {Bardin}, \citenamefont {Barends}, \citenamefont {Biswas},
  \citenamefont {Boixo}, \citenamefont {Brandao}, \citenamefont {Buell} \emph
  {et~al.}}]{arute2019quantum}%
  \BibitemOpen
  \bibfield  {author} {\bibinfo {author} {\bibfnamefont {F.}~\bibnamefont
  {Arute}}, \bibinfo {author} {\bibfnamefont {K.}~\bibnamefont {Arya}},
  \bibinfo {author} {\bibfnamefont {R.}~\bibnamefont {Babbush}}, \bibinfo
  {author} {\bibfnamefont {D.}~\bibnamefont {Bacon}}, \bibinfo {author}
  {\bibfnamefont {J.~C.}\ \bibnamefont {Bardin}}, \bibinfo {author}
  {\bibfnamefont {R.}~\bibnamefont {Barends}}, \bibinfo {author} {\bibfnamefont
  {R.}~\bibnamefont {Biswas}}, \bibinfo {author} {\bibfnamefont
  {S.}~\bibnamefont {Boixo}}, \bibinfo {author} {\bibfnamefont {F.~G.}\
  \bibnamefont {Brandao}}, \bibinfo {author} {\bibfnamefont {D.~A.}\
  \bibnamefont {Buell}}, \emph {et~al.},\ }\bibfield  {title} {\bibinfo {title}
  {Quantum supremacy using a programmable superconducting processor},\ }\href
  {https://doi.org/10.1038/s41586-019-1666-5} {\bibfield  {journal} {\bibinfo
  {journal} {Nature}\ }\textbf {\bibinfo {volume} {574}},\ \bibinfo {pages}
  {505} (\bibinfo {year} {2019})}\BibitemShut {NoStop}%
\bibitem [{\citenamefont {Nakagawa}\ \emph {et~al.}(2024)\citenamefont
  {Nakagawa}, \citenamefont {Kamoshita}, \citenamefont {Mizukami},
  \citenamefont {Sudo},\ and\ \citenamefont {Ohnishi}}]{nakagawa2023adapt}%
  \BibitemOpen
  \bibfield  {author} {\bibinfo {author} {\bibfnamefont {Y.~O.}\ \bibnamefont
  {Nakagawa}}, \bibinfo {author} {\bibfnamefont {M.}~\bibnamefont {Kamoshita}},
  \bibinfo {author} {\bibfnamefont {W.}~\bibnamefont {Mizukami}}, \bibinfo
  {author} {\bibfnamefont {S.}~\bibnamefont {Sudo}},\ and\ \bibinfo {author}
  {\bibfnamefont {Y.-y.}\ \bibnamefont {Ohnishi}},\ }\bibfield  {title}
  {\bibinfo {title} {{ADAPT-QSCI: Adaptive Construction of an Input State for
  Quantum-Selected Configuration Interaction}},\ }\href
  {https://doi.org/10.1021/acs.jctc.4c00846} {\bibfield  {journal} {\bibinfo
  {journal} {Journal of Chemical Theory and Computation}\ }\textbf {\bibinfo
  {volume} {20}},\ \bibinfo {pages} {10817} (\bibinfo {year}
  {2024})}\BibitemShut {NoStop}%
\bibitem [{\citenamefont {Grimsley}\ \emph {et~al.}(2019)\citenamefont
  {Grimsley}, \citenamefont {Economou}, \citenamefont {Barnes},\ and\
  \citenamefont {Mayhall}}]{grimsley2019adaptive}%
  \BibitemOpen
  \bibfield  {author} {\bibinfo {author} {\bibfnamefont {H.~R.}\ \bibnamefont
  {Grimsley}}, \bibinfo {author} {\bibfnamefont {S.~E.}\ \bibnamefont
  {Economou}}, \bibinfo {author} {\bibfnamefont {E.}~\bibnamefont {Barnes}},\
  and\ \bibinfo {author} {\bibfnamefont {N.~J.}\ \bibnamefont {Mayhall}},\
  }\bibfield  {title} {\bibinfo {title} {An adaptive variational algorithm for
  exact molecular simulations on a quantum computer},\ }\href@noop {}
  {\bibfield  {journal} {\bibinfo  {journal} {Nature communications}\ }\textbf
  {\bibinfo {volume} {10}},\ \bibinfo {pages} {3007} (\bibinfo {year}
  {2019})}\BibitemShut {NoStop}%
\bibitem [{\citenamefont {Robledo-Moreno}\ \emph {et~al.}(2024)\citenamefont
  {Robledo-Moreno}, \citenamefont {Motta}, \citenamefont {Haas}, \citenamefont
  {Javadi-Abhari}, \citenamefont {Jurcevic}, \citenamefont {Kirby},
  \citenamefont {Martiel}, \citenamefont {Sharma}, \citenamefont {Sharma},
  \citenamefont {Shirakawa}, \citenamefont {Sitdikov}, \citenamefont {Sun},
  \citenamefont {Sung}, \citenamefont {Takita}, \citenamefont {Tran},
  \citenamefont {Yunoki},\ and\ \citenamefont {Mezzacapo}}]{robledomoreno2024}%
  \BibitemOpen
  \bibfield  {author} {\bibinfo {author} {\bibfnamefont {J.}~\bibnamefont
  {Robledo-Moreno}}, \bibinfo {author} {\bibfnamefont {M.}~\bibnamefont
  {Motta}}, \bibinfo {author} {\bibfnamefont {H.}~\bibnamefont {Haas}},
  \bibinfo {author} {\bibfnamefont {A.}~\bibnamefont {Javadi-Abhari}}, \bibinfo
  {author} {\bibfnamefont {P.}~\bibnamefont {Jurcevic}}, \bibinfo {author}
  {\bibfnamefont {W.}~\bibnamefont {Kirby}}, \bibinfo {author} {\bibfnamefont
  {S.}~\bibnamefont {Martiel}}, \bibinfo {author} {\bibfnamefont
  {K.}~\bibnamefont {Sharma}}, \bibinfo {author} {\bibfnamefont
  {S.}~\bibnamefont {Sharma}}, \bibinfo {author} {\bibfnamefont
  {T.}~\bibnamefont {Shirakawa}}, \bibinfo {author} {\bibfnamefont
  {I.}~\bibnamefont {Sitdikov}}, \bibinfo {author} {\bibfnamefont {R.-Y.}\
  \bibnamefont {Sun}}, \bibinfo {author} {\bibfnamefont {K.~J.}\ \bibnamefont
  {Sung}}, \bibinfo {author} {\bibfnamefont {M.}~\bibnamefont {Takita}},
  \bibinfo {author} {\bibfnamefont {M.~C.}\ \bibnamefont {Tran}}, \bibinfo
  {author} {\bibfnamefont {S.}~\bibnamefont {Yunoki}},\ and\ \bibinfo {author}
  {\bibfnamefont {A.}~\bibnamefont {Mezzacapo}},\ }\href
  {https://arxiv.org/abs/2405.05068} {\bibinfo {title} {Chemistry beyond exact
  solutions on a quantum-centric supercomputer}} (\bibinfo {year} {2024}),\
  \Eprint {https://arxiv.org/abs/2405.05068} {arXiv:2405.05068 [quant-ph]}
  \BibitemShut {NoStop}%
\bibitem [{\citenamefont {Sugisaki}\ \emph {et~al.}(2024)\citenamefont
  {Sugisaki}, \citenamefont {Kanno}, \citenamefont {Itoko}, \citenamefont
  {Sakuma},\ and\ \citenamefont {Yamamoto}}]{sugisaki2024}%
  \BibitemOpen
  \bibfield  {author} {\bibinfo {author} {\bibfnamefont {K.}~\bibnamefont
  {Sugisaki}}, \bibinfo {author} {\bibfnamefont {S.}~\bibnamefont {Kanno}},
  \bibinfo {author} {\bibfnamefont {T.}~\bibnamefont {Itoko}}, \bibinfo
  {author} {\bibfnamefont {R.}~\bibnamefont {Sakuma}},\ and\ \bibinfo {author}
  {\bibfnamefont {N.}~\bibnamefont {Yamamoto}},\ }\href
  {https://arxiv.org/abs/2412.07218} {\bibinfo {title} {Hamiltonian
  simulation-based quantum-selected configuration interaction for large-scale
  electronic structure calculations with a quantum computer}} (\bibinfo {year}
  {2024}),\ \Eprint {https://arxiv.org/abs/2412.07218} {arXiv:2412.07218
  [quant-ph]} \BibitemShut {NoStop}%
\bibitem [{\citenamefont {Jordan}\ and\ \citenamefont
  {Wigner}(1928)}]{Jordan1928}%
  \BibitemOpen
  \bibfield  {author} {\bibinfo {author} {\bibfnamefont {P.}~\bibnamefont
  {Jordan}}\ and\ \bibinfo {author} {\bibfnamefont {E.}~\bibnamefont
  {Wigner}},\ }\bibfield  {title} {\bibinfo {title} {{\"U}ber das paulische
  {\"a}quivalenzverbot},\ }\href {https://doi.org/10.1007/BF01331938}
  {\bibfield  {journal} {\bibinfo  {journal} {Zeitschrift f{\"u}r Physik}\
  }\textbf {\bibinfo {volume} {47}},\ \bibinfo {pages} {631} (\bibinfo {year}
  {1928})}\BibitemShut {NoStop}%
\bibitem [{\citenamefont {Bravyi}\ and\ \citenamefont
  {Kitaev}(2002)}]{Bravyi2002}%
  \BibitemOpen
  \bibfield  {author} {\bibinfo {author} {\bibfnamefont {S.~B.}\ \bibnamefont
  {Bravyi}}\ and\ \bibinfo {author} {\bibfnamefont {A.~Y.}\ \bibnamefont
  {Kitaev}},\ }\bibfield  {title} {\bibinfo {title} {Fermionic quantum
  computation},\ }\href
  {https://doi.org/https://doi.org/10.1006/aphy.2002.6254} {\bibfield
  {journal} {\bibinfo  {journal} {Annals of Physics}\ }\textbf {\bibinfo
  {volume} {298}},\ \bibinfo {pages} {210} (\bibinfo {year}
  {2002})}\BibitemShut {NoStop}%
\bibitem [{\citenamefont {Seeley}\ \emph {et~al.}(2012)\citenamefont {Seeley},
  \citenamefont {Richard},\ and\ \citenamefont {Love}}]{Seeley2012}%
  \BibitemOpen
  \bibfield  {author} {\bibinfo {author} {\bibfnamefont {J.~T.}\ \bibnamefont
  {Seeley}}, \bibinfo {author} {\bibfnamefont {M.~J.}\ \bibnamefont
  {Richard}},\ and\ \bibinfo {author} {\bibfnamefont {P.~J.}\ \bibnamefont
  {Love}},\ }\bibfield  {title} {\bibinfo {title} {The bravyi-kitaev
  transformation for quantum computation of electronic structure},\ }\href
  {https://doi.org/10.1063/1.4768229} {\bibfield  {journal} {\bibinfo
  {journal} {The Journal of Chemical Physics}\ }\textbf {\bibinfo {volume}
  {137}},\ \bibinfo {pages} {224109} (\bibinfo {year} {2012})}\BibitemShut
  {NoStop}%
\bibitem [{\citenamefont {Helgaker}\ \emph {et~al.}(2000)\citenamefont
  {Helgaker}, \citenamefont {J{\o}rgensen},\ and\ \citenamefont
  {Olsen}}]{helgaker2014molecular}%
  \BibitemOpen
  \bibfield  {author} {\bibinfo {author} {\bibfnamefont {T.}~\bibnamefont
  {Helgaker}}, \bibinfo {author} {\bibfnamefont {P.}~\bibnamefont
  {J{\o}rgensen}},\ and\ \bibinfo {author} {\bibfnamefont {J.}~\bibnamefont
  {Olsen}},\ }\href@noop {} {\emph {\bibinfo {title} {Molecular
  Electronic-Structure Theory}}}\ (\bibinfo  {publisher} {John Wiley \& Sons},\
  \bibinfo {year} {2000})\BibitemShut {NoStop}%
\bibitem [{\citenamefont {Bender}\ and\ \citenamefont
  {Davidson}(1969)}]{bender1969pr}%
  \BibitemOpen
  \bibfield  {author} {\bibinfo {author} {\bibfnamefont {C.~F.}\ \bibnamefont
  {Bender}}\ and\ \bibinfo {author} {\bibfnamefont {E.~R.}\ \bibnamefont
  {Davidson}},\ }\bibfield  {title} {\bibinfo {title} {Studies in configuration
  interaction: The first-row diatomic hydrides},\ }\href@noop {} {\bibfield
  {journal} {\bibinfo  {journal} {Physical Review}\ }\textbf {\bibinfo {volume}
  {183}},\ \bibinfo {pages} {23} (\bibinfo {year} {1969})}\BibitemShut
  {NoStop}%
\bibitem [{\citenamefont {Whitten}\ and\ \citenamefont
  {Hackmeyer}(1969)}]{whitten1969jcp}%
  \BibitemOpen
  \bibfield  {author} {\bibinfo {author} {\bibfnamefont {J.}~\bibnamefont
  {Whitten}}\ and\ \bibinfo {author} {\bibfnamefont {M.}~\bibnamefont
  {Hackmeyer}},\ }\bibfield  {title} {\bibinfo {title} {Configuration
  interaction studies of ground and excited states of polyatomic molecules. i.
  the ci formulation and studies of formaldehyde},\ }\href@noop {} {\bibfield
  {journal} {\bibinfo  {journal} {The Journal of Chemical Physics}\ }\textbf
  {\bibinfo {volume} {51}},\ \bibinfo {pages} {5584} (\bibinfo {year}
  {1969})}\BibitemShut {NoStop}%
\bibitem [{\citenamefont {Huron}\ \emph {et~al.}(1973)\citenamefont {Huron},
  \citenamefont {Malrieu},\ and\ \citenamefont {Rancurel}}]{huron1973jcp}%
  \BibitemOpen
  \bibfield  {author} {\bibinfo {author} {\bibfnamefont {B.}~\bibnamefont
  {Huron}}, \bibinfo {author} {\bibfnamefont {J.}~\bibnamefont {Malrieu}},\
  and\ \bibinfo {author} {\bibfnamefont {P.}~\bibnamefont {Rancurel}},\
  }\bibfield  {title} {\bibinfo {title} {Iterative perturbation calculations of
  ground and excited state energies from multiconfigurational zeroth-order
  wavefunctions},\ }\href@noop {} {\bibfield  {journal} {\bibinfo  {journal}
  {The Journal of Chemical Physics}\ }\textbf {\bibinfo {volume} {58}},\
  \bibinfo {pages} {5745} (\bibinfo {year} {1973})}\BibitemShut {NoStop}%
\bibitem [{\citenamefont {Buenker}\ and\ \citenamefont
  {Peyerimhoff}(1974)}]{buenker1974tca}%
  \BibitemOpen
  \bibfield  {author} {\bibinfo {author} {\bibfnamefont {R.~J.}\ \bibnamefont
  {Buenker}}\ and\ \bibinfo {author} {\bibfnamefont {S.~D.}\ \bibnamefont
  {Peyerimhoff}},\ }\bibfield  {title} {\bibinfo {title} {{Individualized
  configuration selection in CI calculations with subsequent energy
  extrapolation}},\ }\href@noop {} {\bibfield  {journal} {\bibinfo  {journal}
  {Theoretica chimica acta}\ }\textbf {\bibinfo {volume} {35}},\ \bibinfo
  {pages} {33} (\bibinfo {year} {1974})}\BibitemShut {NoStop}%
\bibitem [{\citenamefont {Buenker}\ and\ \citenamefont
  {Peyerimhoff}(1975)}]{buenker1975tca}%
  \BibitemOpen
  \bibfield  {author} {\bibinfo {author} {\bibfnamefont {R.~J.}\ \bibnamefont
  {Buenker}}\ and\ \bibinfo {author} {\bibfnamefont {S.~D.}\ \bibnamefont
  {Peyerimhoff}},\ }\bibfield  {title} {\bibinfo {title} {{Energy extrapolation
  in CI calculations}},\ }\href@noop {} {\bibfield  {journal} {\bibinfo
  {journal} {Theoretica chimica acta}\ }\textbf {\bibinfo {volume} {39}},\
  \bibinfo {pages} {217} (\bibinfo {year} {1975})}\BibitemShut {NoStop}%
\bibitem [{\citenamefont {Nakatsuji}(1983)}]{nakatsuji1983cluster}%
  \BibitemOpen
  \bibfield  {author} {\bibinfo {author} {\bibfnamefont {H.}~\bibnamefont
  {Nakatsuji}},\ }\bibfield  {title} {\bibinfo {title} {Cluster expansion of
  the wavefunction, valence and rydberg excitations, ionizations, and
  inner-valence ionizations of {CO$_2$} and {N$_2$O} studied by the sac and sac
  {CI} theories},\ }\href@noop {} {\bibfield  {journal} {\bibinfo  {journal}
  {Chemical Physics}\ }\textbf {\bibinfo {volume} {75}},\ \bibinfo {pages}
  {425} (\bibinfo {year} {1983})}\BibitemShut {NoStop}%
\bibitem [{\citenamefont {Cimiraglia}\ and\ \citenamefont
  {Persico}(1987)}]{cimiraglia1987jcc}%
  \BibitemOpen
  \bibfield  {author} {\bibinfo {author} {\bibfnamefont {R.}~\bibnamefont
  {Cimiraglia}}\ and\ \bibinfo {author} {\bibfnamefont {M.}~\bibnamefont
  {Persico}},\ }\bibfield  {title} {\bibinfo {title} {{Recent Advances in
  Multireference Second Order Perturbation CI: The CIPSI Method Revisited}},\
  }\href@noop {} {\bibfield  {journal} {\bibinfo  {journal} {J. Comput. Chem.}\
  }\textbf {\bibinfo {volume} {8}},\ \bibinfo {pages} {39} (\bibinfo {year}
  {1987})}\BibitemShut {NoStop}%
\bibitem [{\citenamefont {Harrison}(1991)}]{harrison1991jcp}%
  \BibitemOpen
  \bibfield  {author} {\bibinfo {author} {\bibfnamefont {R.~J.}\ \bibnamefont
  {Harrison}},\ }\bibfield  {title} {\bibinfo {title} {{Approximating Full
  Configuration Interaction with Selected Configuration Interaction and
  Perturbation Theory}},\ }\href@noop {} {\bibfield  {journal} {\bibinfo
  {journal} {J. Chem. Phys.}\ }\textbf {\bibinfo {volume} {94}},\ \bibinfo
  {pages} {5021} (\bibinfo {year} {1991})}\BibitemShut {NoStop}%
\bibitem [{\citenamefont {Greer}(1995)}]{greer1995jcp}%
  \BibitemOpen
  \bibfield  {author} {\bibinfo {author} {\bibfnamefont {J.}~\bibnamefont
  {Greer}},\ }\bibfield  {title} {\bibinfo {title} {{Estimating full
  configuration interaction limits from a Monte Carlo selection of the
  expansion space}},\ }\href@noop {} {\bibfield  {journal} {\bibinfo  {journal}
  {The Journal of Chemical Physics}\ }\textbf {\bibinfo {volume} {103}},\
  \bibinfo {pages} {1821} (\bibinfo {year} {1995})}\BibitemShut {NoStop}%
\bibitem [{\citenamefont {Greer}(1998)}]{greer1998jcpss}%
  \BibitemOpen
  \bibfield  {author} {\bibinfo {author} {\bibfnamefont {J.}~\bibnamefont
  {Greer}},\ }\bibfield  {title} {\bibinfo {title} {Monte carlo configuration
  interaction},\ }\href@noop {} {\bibfield  {journal} {\bibinfo  {journal}
  {Journal of Computational Physics}\ }\textbf {\bibinfo {volume} {146}},\
  \bibinfo {pages} {181} (\bibinfo {year} {1998})}\BibitemShut {NoStop}%
\bibitem [{\citenamefont {Evangelista}(2014)}]{evangelista2014jcp}%
  \BibitemOpen
  \bibfield  {author} {\bibinfo {author} {\bibfnamefont {F.~A.}\ \bibnamefont
  {Evangelista}},\ }\bibfield  {title} {\bibinfo {title} {Adaptive
  multiconfigurational wave functions},\ }\href@noop {} {\bibfield  {journal}
  {\bibinfo  {journal} {J. Chem. Phys.}\ }\textbf {\bibinfo {volume} {140}},\
  \bibinfo {pages} {124114} (\bibinfo {year} {2014})}\BibitemShut {NoStop}%
\bibitem [{\citenamefont {Holmes}\ \emph
  {et~al.}(2016{\natexlab{a}})\citenamefont {Holmes}, \citenamefont
  {Changlani},\ and\ \citenamefont {Umrigar}}]{holmes2016jctc}%
  \BibitemOpen
  \bibfield  {author} {\bibinfo {author} {\bibfnamefont {A.~A.}\ \bibnamefont
  {Holmes}}, \bibinfo {author} {\bibfnamefont {H.~J.}\ \bibnamefont
  {Changlani}},\ and\ \bibinfo {author} {\bibfnamefont {C.}~\bibnamefont
  {Umrigar}},\ }\bibfield  {title} {\bibinfo {title} {{Efficient heat-bath
  sampling in Fock space}},\ }\href@noop {} {\bibfield  {journal} {\bibinfo
  {journal} {J. Chem. Theory Comput.}\ }\textbf {\bibinfo {volume} {12}},\
  \bibinfo {pages} {1561} (\bibinfo {year} {2016}{\natexlab{a}})}\BibitemShut
  {NoStop}%
\bibitem [{\citenamefont {Schriber}\ and\ \citenamefont
  {Evangelista}(2016)}]{schriber2016jcp}%
  \BibitemOpen
  \bibfield  {author} {\bibinfo {author} {\bibfnamefont {J.~B.}\ \bibnamefont
  {Schriber}}\ and\ \bibinfo {author} {\bibfnamefont {F.~A.}\ \bibnamefont
  {Evangelista}},\ }\bibfield  {title} {\bibinfo {title} {Communication: An
  adaptive configuration interaction approach for strongly correlated electrons
  with tunable accuracy},\ }\href@noop {} {\bibfield  {journal} {\bibinfo
  {journal} {J. Chem. Phys.}\ }\textbf {\bibinfo {volume} {144}},\ \bibinfo
  {pages} {161106} (\bibinfo {year} {2016})}\BibitemShut {NoStop}%
\bibitem [{\citenamefont {Holmes}\ \emph
  {et~al.}(2016{\natexlab{b}})\citenamefont {Holmes}, \citenamefont {Tubman},\
  and\ \citenamefont {Umrigar}}]{holmes2016jctc2}%
  \BibitemOpen
  \bibfield  {author} {\bibinfo {author} {\bibfnamefont {A.~A.}\ \bibnamefont
  {Holmes}}, \bibinfo {author} {\bibfnamefont {N.~M.}\ \bibnamefont {Tubman}},\
  and\ \bibinfo {author} {\bibfnamefont {C.}~\bibnamefont {Umrigar}},\
  }\bibfield  {title} {\bibinfo {title} {Heat-bath configuration interaction:
  An efficient selected configuration interaction algorithm inspired by
  heat-bath sampling},\ }\href@noop {} {\bibfield  {journal} {\bibinfo
  {journal} {Journal of chemical theory and computation}\ }\textbf {\bibinfo
  {volume} {12}},\ \bibinfo {pages} {3674} (\bibinfo {year}
  {2016}{\natexlab{b}})}\BibitemShut {NoStop}%
\bibitem [{\citenamefont {Tubman}\ \emph {et~al.}(2016)\citenamefont {Tubman},
  \citenamefont {Lee}, \citenamefont {Takeshita}, \citenamefont {Head-Gordon},\
  and\ \citenamefont {Whaley}}]{tubman2016deterministic}%
  \BibitemOpen
  \bibfield  {author} {\bibinfo {author} {\bibfnamefont {N.~M.}\ \bibnamefont
  {Tubman}}, \bibinfo {author} {\bibfnamefont {J.}~\bibnamefont {Lee}},
  \bibinfo {author} {\bibfnamefont {T.~Y.}\ \bibnamefont {Takeshita}}, \bibinfo
  {author} {\bibfnamefont {M.}~\bibnamefont {Head-Gordon}},\ and\ \bibinfo
  {author} {\bibfnamefont {K.~B.}\ \bibnamefont {Whaley}},\ }\bibfield  {title}
  {\bibinfo {title} {{A deterministic alternative to the full configuration
  interaction quantum Monte Carlo method}},\ }\href@noop {} {\bibfield
  {journal} {\bibinfo  {journal} {The Journal of chemical physics}\ }\textbf
  {\bibinfo {volume} {145}},\ \bibinfo {pages} {044112} (\bibinfo {year}
  {2016})}\BibitemShut {NoStop}%
\bibitem [{\citenamefont {Ohtsuka}\ and\ \citenamefont
  {Hasegawa}(2017)}]{ohtsuka2017jcp}%
  \BibitemOpen
  \bibfield  {author} {\bibinfo {author} {\bibfnamefont {Y.}~\bibnamefont
  {Ohtsuka}}\ and\ \bibinfo {author} {\bibfnamefont {J.}~\bibnamefont
  {Hasegawa}},\ }\bibfield  {title} {\bibinfo {title} {Selected configuration
  interaction using sampled first-order corrections to wave functions},\
  }\href@noop {} {\bibfield  {journal} {\bibinfo  {journal} {J. Chem. Phys.}\
  }\textbf {\bibinfo {volume} {147}},\ \bibinfo {pages} {034102} (\bibinfo
  {year} {2017})}\BibitemShut {NoStop}%
\bibitem [{\citenamefont {Schriber}\ and\ \citenamefont
  {Evangelista}(2017)}]{schriber2017jctc}%
  \BibitemOpen
  \bibfield  {author} {\bibinfo {author} {\bibfnamefont {J.~B.}\ \bibnamefont
  {Schriber}}\ and\ \bibinfo {author} {\bibfnamefont {F.~A.}\ \bibnamefont
  {Evangelista}},\ }\bibfield  {title} {\bibinfo {title} {Adaptive
  configuration interaction for computing challenging electronic excited states
  with tunable accuracy},\ }\href@noop {} {\bibfield  {journal} {\bibinfo
  {journal} {J. Chem. Theory Comput.}\ }\textbf {\bibinfo {volume} {13}},\
  \bibinfo {pages} {5354} (\bibinfo {year} {2017})}\BibitemShut {NoStop}%
\bibitem [{\citenamefont {Sharma}\ \emph {et~al.}(2017)\citenamefont {Sharma},
  \citenamefont {Holmes}, \citenamefont {Jeanmairet}, \citenamefont {Alavi},\
  and\ \citenamefont {Umrigar}}]{sharma2017semistochastic}%
  \BibitemOpen
  \bibfield  {author} {\bibinfo {author} {\bibfnamefont {S.}~\bibnamefont
  {Sharma}}, \bibinfo {author} {\bibfnamefont {A.~A.}\ \bibnamefont {Holmes}},
  \bibinfo {author} {\bibfnamefont {G.}~\bibnamefont {Jeanmairet}}, \bibinfo
  {author} {\bibfnamefont {A.}~\bibnamefont {Alavi}},\ and\ \bibinfo {author}
  {\bibfnamefont {C.~J.}\ \bibnamefont {Umrigar}},\ }\bibfield  {title}
  {\bibinfo {title} {Semistochastic heat-bath configuration interaction method:
  Selected configuration interaction with semistochastic perturbation theory},\
  }\href@noop {} {\bibfield  {journal} {\bibinfo  {journal} {Journal of
  chemical theory and computation}\ }\textbf {\bibinfo {volume} {13}},\
  \bibinfo {pages} {1595} (\bibinfo {year} {2017})}\BibitemShut {NoStop}%
\bibitem [{\citenamefont {Chakraborty}\ \emph {et~al.}(2018)\citenamefont
  {Chakraborty}, \citenamefont {Ghosh},\ and\ \citenamefont
  {Ghosh}}]{chakraborty2018ijqc}%
  \BibitemOpen
  \bibfield  {author} {\bibinfo {author} {\bibfnamefont {R.}~\bibnamefont
  {Chakraborty}}, \bibinfo {author} {\bibfnamefont {P.}~\bibnamefont {Ghosh}},\
  and\ \bibinfo {author} {\bibfnamefont {D.}~\bibnamefont {Ghosh}},\ }\bibfield
   {title} {\bibinfo {title} {Evolutionary algorithm based configuration
  interaction approach},\ }\href@noop {} {\bibfield  {journal} {\bibinfo
  {journal} {International Journal of Quantum Chemistry}\ }\textbf {\bibinfo
  {volume} {118}},\ \bibinfo {pages} {e25509} (\bibinfo {year}
  {2018})}\BibitemShut {NoStop}%
\bibitem [{\citenamefont {Coe}(2018)}]{coe2018jctc}%
  \BibitemOpen
  \bibfield  {author} {\bibinfo {author} {\bibfnamefont {J.~P.}\ \bibnamefont
  {Coe}},\ }\bibfield  {title} {\bibinfo {title} {Machine learning
  configuration interaction},\ }\href@noop {} {\bibfield  {journal} {\bibinfo
  {journal} {Journal of chemical theory and computation}\ }\textbf {\bibinfo
  {volume} {14}},\ \bibinfo {pages} {5739} (\bibinfo {year}
  {2018})}\BibitemShut {NoStop}%
\bibitem [{\citenamefont {Coe}(2019)}]{coe2019jctc}%
  \BibitemOpen
  \bibfield  {author} {\bibinfo {author} {\bibfnamefont {J.~P.}\ \bibnamefont
  {Coe}},\ }\bibfield  {title} {\bibinfo {title} {Machine learning
  configuration interaction for ab initio potential energy curves},\
  }\href@noop {} {\bibfield  {journal} {\bibinfo  {journal} {Journal of
  chemical theory and computation}\ }\textbf {\bibinfo {volume} {15}},\
  \bibinfo {pages} {6179} (\bibinfo {year} {2019})}\BibitemShut {NoStop}%
\bibitem [{\citenamefont {Abraham}\ and\ \citenamefont
  {Mayhall}(2020)}]{abraham202jctc}%
  \BibitemOpen
  \bibfield  {author} {\bibinfo {author} {\bibfnamefont {V.}~\bibnamefont
  {Abraham}}\ and\ \bibinfo {author} {\bibfnamefont {N.~J.}\ \bibnamefont
  {Mayhall}},\ }\bibfield  {title} {\bibinfo {title} {Selected configuration
  interaction in a basis of cluster state tensor products},\ }\href@noop {}
  {\bibfield  {journal} {\bibinfo  {journal} {Journal of Chemical Theory and
  Computation}\ }\textbf {\bibinfo {volume} {16}},\ \bibinfo {pages} {6098}
  (\bibinfo {year} {2020})}\BibitemShut {NoStop}%
\bibitem [{\citenamefont {Tubman}\ \emph {et~al.}(2020)\citenamefont {Tubman},
  \citenamefont {Freeman}, \citenamefont {Levine}, \citenamefont {Hait},
  \citenamefont {Head-Gordon},\ and\ \citenamefont
  {Whaley}}]{tubman2020modern}%
  \BibitemOpen
  \bibfield  {author} {\bibinfo {author} {\bibfnamefont {N.~M.}\ \bibnamefont
  {Tubman}}, \bibinfo {author} {\bibfnamefont {C.~D.}\ \bibnamefont {Freeman}},
  \bibinfo {author} {\bibfnamefont {D.~S.}\ \bibnamefont {Levine}}, \bibinfo
  {author} {\bibfnamefont {D.}~\bibnamefont {Hait}}, \bibinfo {author}
  {\bibfnamefont {M.}~\bibnamefont {Head-Gordon}},\ and\ \bibinfo {author}
  {\bibfnamefont {K.~B.}\ \bibnamefont {Whaley}},\ }\bibfield  {title}
  {\bibinfo {title} {{Modern approaches to exact diagonalization and selected
  configuration interaction with the adaptive sampling CI method}},\
  }\href@noop {} {\bibfield  {journal} {\bibinfo  {journal} {Journal of
  chemical theory and computation}\ }\textbf {\bibinfo {volume} {16}},\
  \bibinfo {pages} {2139} (\bibinfo {year} {2020})}\BibitemShut {NoStop}%
\bibitem [{\citenamefont {Zhang}\ \emph {et~al.}(2020)\citenamefont {Zhang},
  \citenamefont {Liu},\ and\ \citenamefont {Hoffmann}}]{zhang2020jctc}%
  \BibitemOpen
  \bibfield  {author} {\bibinfo {author} {\bibfnamefont {N.}~\bibnamefont
  {Zhang}}, \bibinfo {author} {\bibfnamefont {W.}~\bibnamefont {Liu}},\ and\
  \bibinfo {author} {\bibfnamefont {M.~R.}\ \bibnamefont {Hoffmann}},\
  }\bibfield  {title} {\bibinfo {title} {Iterative configuration interaction
  with selection},\ }\href@noop {} {\bibfield  {journal} {\bibinfo  {journal}
  {Journal of Chemical Theory and Computation}\ }\textbf {\bibinfo {volume}
  {16}},\ \bibinfo {pages} {2296} (\bibinfo {year} {2020})}\BibitemShut
  {NoStop}%
\bibitem [{\citenamefont {Zhang}\ \emph {et~al.}(2021)\citenamefont {Zhang},
  \citenamefont {Liu},\ and\ \citenamefont {Hoffmann}}]{zhang2021jctc}%
  \BibitemOpen
  \bibfield  {author} {\bibinfo {author} {\bibfnamefont {N.}~\bibnamefont
  {Zhang}}, \bibinfo {author} {\bibfnamefont {W.}~\bibnamefont {Liu}},\ and\
  \bibinfo {author} {\bibfnamefont {M.~R.}\ \bibnamefont {Hoffmann}},\
  }\bibfield  {title} {\bibinfo {title} {{Further development of iCIPT2 for
  strongly correlated electrons}},\ }\href@noop {} {\bibfield  {journal}
  {\bibinfo  {journal} {J. Chem. Theory Comput.}\ }\textbf {\bibinfo {volume}
  {17}},\ \bibinfo {pages} {949} (\bibinfo {year} {2021})}\BibitemShut
  {NoStop}%
\bibitem [{\citenamefont {Chilkuri}\ and\ \citenamefont
  {Neese}(2021{\natexlab{a}})}]{chilkuri2021jcc}%
  \BibitemOpen
  \bibfield  {author} {\bibinfo {author} {\bibfnamefont {V.~G.}\ \bibnamefont
  {Chilkuri}}\ and\ \bibinfo {author} {\bibfnamefont {F.}~\bibnamefont
  {Neese}},\ }\bibfield  {title} {\bibinfo {title} {{Comparison of
  many-particle representations for selected-CI I: A tree based approach}},\
  }\href@noop {} {\bibfield  {journal} {\bibinfo  {journal} {Journal of
  Computational Chemistry}\ }\textbf {\bibinfo {volume} {42}},\ \bibinfo
  {pages} {982} (\bibinfo {year} {2021}{\natexlab{a}})}\BibitemShut {NoStop}%
\bibitem [{\citenamefont {Chilkuri}\ and\ \citenamefont
  {Neese}(2021{\natexlab{b}})}]{chilkuri2021jctc}%
  \BibitemOpen
  \bibfield  {author} {\bibinfo {author} {\bibfnamefont {V.~G.}\ \bibnamefont
  {Chilkuri}}\ and\ \bibinfo {author} {\bibfnamefont {F.}~\bibnamefont
  {Neese}},\ }\bibfield  {title} {\bibinfo {title} {{Comparison of
  many-particle representations for selected configuration interaction: II.
  Numerical benchmark calculations}},\ }\href@noop {} {\bibfield  {journal}
  {\bibinfo  {journal} {Journal of Chemical Theory and Computation}\ }\textbf
  {\bibinfo {volume} {17}},\ \bibinfo {pages} {2868} (\bibinfo {year}
  {2021}{\natexlab{b}})}\BibitemShut {NoStop}%
\bibitem [{\citenamefont {Goings}\ \emph {et~al.}(2021)\citenamefont {Goings},
  \citenamefont {Hu}, \citenamefont {Yang},\ and\ \citenamefont
  {Li}}]{goings2021jctc}%
  \BibitemOpen
  \bibfield  {author} {\bibinfo {author} {\bibfnamefont {J.~J.}\ \bibnamefont
  {Goings}}, \bibinfo {author} {\bibfnamefont {H.}~\bibnamefont {Hu}}, \bibinfo
  {author} {\bibfnamefont {C.}~\bibnamefont {Yang}},\ and\ \bibinfo {author}
  {\bibfnamefont {X.}~\bibnamefont {Li}},\ }\bibfield  {title} {\bibinfo
  {title} {Reinforcement learning configuration interaction},\ }\href@noop {}
  {\bibfield  {journal} {\bibinfo  {journal} {Journal of chemical theory and
  computation}\ }\textbf {\bibinfo {volume} {17}},\ \bibinfo {pages} {5482}
  (\bibinfo {year} {2021})}\BibitemShut {NoStop}%
\bibitem [{\citenamefont {Pineda~Flores}(2021)}]{pineda2021jctc}%
  \BibitemOpen
  \bibfield  {author} {\bibinfo {author} {\bibfnamefont {S.~D.}\ \bibnamefont
  {Pineda~Flores}},\ }\bibfield  {title} {\bibinfo {title} {Chembot: a machine
  learning approach to selective configuration interaction},\ }\href@noop {}
  {\bibfield  {journal} {\bibinfo  {journal} {Journal of Chemical Theory and
  Computation}\ }\textbf {\bibinfo {volume} {17}},\ \bibinfo {pages} {4028}
  (\bibinfo {year} {2021})}\BibitemShut {NoStop}%
\bibitem [{\citenamefont {Jeong}\ \emph {et~al.}(2021)\citenamefont {Jeong},
  \citenamefont {Gaggioli},\ and\ \citenamefont {Gagliardi}}]{jeong2021jctc}%
  \BibitemOpen
  \bibfield  {author} {\bibinfo {author} {\bibfnamefont {W.}~\bibnamefont
  {Jeong}}, \bibinfo {author} {\bibfnamefont {C.~A.}\ \bibnamefont
  {Gaggioli}},\ and\ \bibinfo {author} {\bibfnamefont {L.}~\bibnamefont
  {Gagliardi}},\ }\bibfield  {title} {\bibinfo {title} {Active learning
  configuration interaction for excited-state calculations of polycyclic
  aromatic hydrocarbons},\ }\href@noop {} {\bibfield  {journal} {\bibinfo
  {journal} {Journal of chemical theory and computation}\ }\textbf {\bibinfo
  {volume} {17}},\ \bibinfo {pages} {7518} (\bibinfo {year}
  {2021})}\BibitemShut {NoStop}%
\bibitem [{\citenamefont {Seth}\ and\ \citenamefont
  {Ghosh}(2023)}]{seth2023jctc}%
  \BibitemOpen
  \bibfield  {author} {\bibinfo {author} {\bibfnamefont {K.}~\bibnamefont
  {Seth}}\ and\ \bibinfo {author} {\bibfnamefont {D.}~\bibnamefont {Ghosh}},\
  }\bibfield  {title} {\bibinfo {title} {{Active Learning Assisted MCCI to
  Target Spin States}},\ }\href {https://doi.org/10.1021/acs.jctc.2c00935}
  {\bibfield  {journal} {\bibinfo  {journal} {Journal of Chemical Theory and
  Computation}\ }\textbf {\bibinfo {volume} {19}},\ \bibinfo {pages} {524}
  (\bibinfo {year} {2023})}\BibitemShut {NoStop}%
\bibitem [{NIS(2022)}]{NIST_CCCBDB}%
  \BibitemOpen
  \href {http://cccbdb.nist.gov/} {\bibinfo {title} {{NIST Computational
  Chemistry Comparison and Benchmark Database}}},\ \bibinfo {howpublished}
  {NIST Standard Reference Database Number 101} (\bibinfo {year} {2022}),\
  \bibinfo {note} {release 22}\BibitemShut {NoStop}%
\bibitem [{\citenamefont {Sun}\ \emph {et~al.}(2018)\citenamefont {Sun},
  \citenamefont {Berkelbach}, \citenamefont {Blunt}, \citenamefont {Booth},
  \citenamefont {Guo}, \citenamefont {Li}, \citenamefont {Liu}, \citenamefont
  {McClain}, \citenamefont {Sayfutyarova}, \citenamefont {Sharma},
  \citenamefont {Wouters},\ and\ \citenamefont {Chan}}]{Sun2018}%
  \BibitemOpen
  \bibfield  {author} {\bibinfo {author} {\bibfnamefont {Q.}~\bibnamefont
  {Sun}}, \bibinfo {author} {\bibfnamefont {T.~C.}\ \bibnamefont {Berkelbach}},
  \bibinfo {author} {\bibfnamefont {N.~S.}\ \bibnamefont {Blunt}}, \bibinfo
  {author} {\bibfnamefont {G.~H.}\ \bibnamefont {Booth}}, \bibinfo {author}
  {\bibfnamefont {S.}~\bibnamefont {Guo}}, \bibinfo {author} {\bibfnamefont
  {Z.}~\bibnamefont {Li}}, \bibinfo {author} {\bibfnamefont {J.}~\bibnamefont
  {Liu}}, \bibinfo {author} {\bibfnamefont {J.~D.}\ \bibnamefont {McClain}},
  \bibinfo {author} {\bibfnamefont {E.~R.}\ \bibnamefont {Sayfutyarova}},
  \bibinfo {author} {\bibfnamefont {S.}~\bibnamefont {Sharma}}, \bibinfo
  {author} {\bibfnamefont {S.}~\bibnamefont {Wouters}},\ and\ \bibinfo {author}
  {\bibfnamefont {G.~K.-L.}\ \bibnamefont {Chan}},\ }\bibfield  {title}
  {\bibinfo {title} {Pyscf: the python-based simulations of chemistry
  framework},\ }\href {https://doi.org/https://doi.org/10.1002/wcms.1340}
  {\bibfield  {journal} {\bibinfo  {journal} {WIREs Computational Molecular
  Science}\ }\textbf {\bibinfo {volume} {8}},\ \bibinfo {pages} {e1340}
  (\bibinfo {year} {2018})}\BibitemShut {NoStop}%
\bibitem [{\citenamefont {Sun}\ \emph {et~al.}(2020)\citenamefont {Sun},
  \citenamefont {Zhang}, \citenamefont {Banerjee}, \citenamefont {Bao},
  \citenamefont {Barbry}, \citenamefont {Blunt}, \citenamefont {Bogdanov},
  \citenamefont {Booth}, \citenamefont {Chen}, \citenamefont {Cui},
  \citenamefont {Eriksen}, \citenamefont {Gao}, \citenamefont {Guo},
  \citenamefont {Hermann}, \citenamefont {Hermes}, \citenamefont {Koh},
  \citenamefont {Koval}, \citenamefont {Lehtola}, \citenamefont {Li},
  \citenamefont {Liu}, \citenamefont {Mardirossian}, \citenamefont {McClain},
  \citenamefont {Motta}, \citenamefont {Mussard}, \citenamefont {Pham},
  \citenamefont {Pulkin}, \citenamefont {Purwanto}, \citenamefont {Robinson},
  \citenamefont {Ronca}, \citenamefont {Sayfutyarova}, \citenamefont
  {Scheurer}, \citenamefont {Schurkus}, \citenamefont {Smith}, \citenamefont
  {Sun}, \citenamefont {Sun}, \citenamefont {Upadhyay}, \citenamefont {Wagner},
  \citenamefont {Wang}, \citenamefont {White}, \citenamefont {Whitfield},
  \citenamefont {Williamson}, \citenamefont {Wouters}, \citenamefont {Yang},
  \citenamefont {Yu}, \citenamefont {Zhu}, \citenamefont {Berkelbach},
  \citenamefont {Sharma}, \citenamefont {Sokolov},\ and\ \citenamefont
  {Chan}}]{Sun2020}%
  \BibitemOpen
  \bibfield  {author} {\bibinfo {author} {\bibfnamefont {Q.}~\bibnamefont
  {Sun}}, \bibinfo {author} {\bibfnamefont {X.}~\bibnamefont {Zhang}}, \bibinfo
  {author} {\bibfnamefont {S.}~\bibnamefont {Banerjee}}, \bibinfo {author}
  {\bibfnamefont {P.}~\bibnamefont {Bao}}, \bibinfo {author} {\bibfnamefont
  {M.}~\bibnamefont {Barbry}}, \bibinfo {author} {\bibfnamefont {N.~S.}\
  \bibnamefont {Blunt}}, \bibinfo {author} {\bibfnamefont {N.~A.}\ \bibnamefont
  {Bogdanov}}, \bibinfo {author} {\bibfnamefont {G.~H.}\ \bibnamefont {Booth}},
  \bibinfo {author} {\bibfnamefont {J.}~\bibnamefont {Chen}}, \bibinfo {author}
  {\bibfnamefont {Z.-H.}\ \bibnamefont {Cui}}, \bibinfo {author} {\bibfnamefont
  {J.~J.}\ \bibnamefont {Eriksen}}, \bibinfo {author} {\bibfnamefont
  {Y.}~\bibnamefont {Gao}}, \bibinfo {author} {\bibfnamefont {S.}~\bibnamefont
  {Guo}}, \bibinfo {author} {\bibfnamefont {J.}~\bibnamefont {Hermann}},
  \bibinfo {author} {\bibfnamefont {M.~R.}\ \bibnamefont {Hermes}}, \bibinfo
  {author} {\bibfnamefont {K.}~\bibnamefont {Koh}}, \bibinfo {author}
  {\bibfnamefont {P.}~\bibnamefont {Koval}}, \bibinfo {author} {\bibfnamefont
  {S.}~\bibnamefont {Lehtola}}, \bibinfo {author} {\bibfnamefont
  {Z.}~\bibnamefont {Li}}, \bibinfo {author} {\bibfnamefont {J.}~\bibnamefont
  {Liu}}, \bibinfo {author} {\bibfnamefont {N.}~\bibnamefont {Mardirossian}},
  \bibinfo {author} {\bibfnamefont {J.~D.}\ \bibnamefont {McClain}}, \bibinfo
  {author} {\bibfnamefont {M.}~\bibnamefont {Motta}}, \bibinfo {author}
  {\bibfnamefont {B.}~\bibnamefont {Mussard}}, \bibinfo {author} {\bibfnamefont
  {H.~Q.}\ \bibnamefont {Pham}}, \bibinfo {author} {\bibfnamefont
  {A.}~\bibnamefont {Pulkin}}, \bibinfo {author} {\bibfnamefont
  {W.}~\bibnamefont {Purwanto}}, \bibinfo {author} {\bibfnamefont {P.~J.}\
  \bibnamefont {Robinson}}, \bibinfo {author} {\bibfnamefont {E.}~\bibnamefont
  {Ronca}}, \bibinfo {author} {\bibfnamefont {E.~R.}\ \bibnamefont
  {Sayfutyarova}}, \bibinfo {author} {\bibfnamefont {M.}~\bibnamefont
  {Scheurer}}, \bibinfo {author} {\bibfnamefont {H.~F.}\ \bibnamefont
  {Schurkus}}, \bibinfo {author} {\bibfnamefont {J.~E.~T.}\ \bibnamefont
  {Smith}}, \bibinfo {author} {\bibfnamefont {C.}~\bibnamefont {Sun}}, \bibinfo
  {author} {\bibfnamefont {S.-N.}\ \bibnamefont {Sun}}, \bibinfo {author}
  {\bibfnamefont {S.}~\bibnamefont {Upadhyay}}, \bibinfo {author}
  {\bibfnamefont {L.~K.}\ \bibnamefont {Wagner}}, \bibinfo {author}
  {\bibfnamefont {X.}~\bibnamefont {Wang}}, \bibinfo {author} {\bibfnamefont
  {A.}~\bibnamefont {White}}, \bibinfo {author} {\bibfnamefont {J.~D.}\
  \bibnamefont {Whitfield}}, \bibinfo {author} {\bibfnamefont {M.~J.}\
  \bibnamefont {Williamson}}, \bibinfo {author} {\bibfnamefont
  {S.}~\bibnamefont {Wouters}}, \bibinfo {author} {\bibfnamefont
  {J.}~\bibnamefont {Yang}}, \bibinfo {author} {\bibfnamefont {J.~M.}\
  \bibnamefont {Yu}}, \bibinfo {author} {\bibfnamefont {T.}~\bibnamefont
  {Zhu}}, \bibinfo {author} {\bibfnamefont {T.~C.}\ \bibnamefont {Berkelbach}},
  \bibinfo {author} {\bibfnamefont {S.}~\bibnamefont {Sharma}}, \bibinfo
  {author} {\bibfnamefont {A.~Y.}\ \bibnamefont {Sokolov}},\ and\ \bibinfo
  {author} {\bibfnamefont {G.~K.-L.}\ \bibnamefont {Chan}},\ }\bibfield
  {title} {\bibinfo {title} {{Recent developments in the PySCF program
  package}},\ }\href {https://doi.org/10.1063/5.0006074} {\bibfield  {journal}
  {\bibinfo  {journal} {The Journal of Chemical Physics}\ }\textbf {\bibinfo
  {volume} {153}},\ \bibinfo {pages} {024109} (\bibinfo {year}
  {2020})}\BibitemShut {NoStop}%
\bibitem [{\citenamefont {McClean}\ \emph {et~al.}(2020)\citenamefont
  {McClean}, \citenamefont {Rubin}, \citenamefont {Sung}, \citenamefont
  {Kivlichan}, \citenamefont {Bonet-Monroig}, \citenamefont {Cao},
  \citenamefont {Dai}, \citenamefont {Fried}, \citenamefont {Gidney},
  \citenamefont {Gimby}, \citenamefont {Gokhale}, \citenamefont {Häner},
  \citenamefont {Hardikar}, \citenamefont {Havlíček}, \citenamefont
  {Higgott}, \citenamefont {Huang}, \citenamefont {Izaac}, \citenamefont
  {Jiang}, \citenamefont {Liu}, \citenamefont {McArdle}, \citenamefont
  {Neeley}, \citenamefont {O’Brien}, \citenamefont {O’Gorman},
  \citenamefont {Ozfidan}, \citenamefont {Radin}, \citenamefont {Romero},
  \citenamefont {Sawaya}, \citenamefont {Senjean}, \citenamefont {Setia},
  \citenamefont {Sim}, \citenamefont {Steiger}, \citenamefont {Steudtner},
  \citenamefont {Sun}, \citenamefont {Sun}, \citenamefont {Wang}, \citenamefont
  {Zhang},\ and\ \citenamefont {Babbush}}]{McClean2020}%
  \BibitemOpen
  \bibfield  {author} {\bibinfo {author} {\bibfnamefont {J.~R.}\ \bibnamefont
  {McClean}}, \bibinfo {author} {\bibfnamefont {N.~C.}\ \bibnamefont {Rubin}},
  \bibinfo {author} {\bibfnamefont {K.~J.}\ \bibnamefont {Sung}}, \bibinfo
  {author} {\bibfnamefont {I.~D.}\ \bibnamefont {Kivlichan}}, \bibinfo {author}
  {\bibfnamefont {X.}~\bibnamefont {Bonet-Monroig}}, \bibinfo {author}
  {\bibfnamefont {Y.}~\bibnamefont {Cao}}, \bibinfo {author} {\bibfnamefont
  {C.}~\bibnamefont {Dai}}, \bibinfo {author} {\bibfnamefont {E.~S.}\
  \bibnamefont {Fried}}, \bibinfo {author} {\bibfnamefont {C.}~\bibnamefont
  {Gidney}}, \bibinfo {author} {\bibfnamefont {B.}~\bibnamefont {Gimby}},
  \bibinfo {author} {\bibfnamefont {P.}~\bibnamefont {Gokhale}}, \bibinfo
  {author} {\bibfnamefont {T.}~\bibnamefont {Häner}}, \bibinfo {author}
  {\bibfnamefont {T.}~\bibnamefont {Hardikar}}, \bibinfo {author}
  {\bibfnamefont {V.}~\bibnamefont {Havlíček}}, \bibinfo {author}
  {\bibfnamefont {O.}~\bibnamefont {Higgott}}, \bibinfo {author} {\bibfnamefont
  {C.}~\bibnamefont {Huang}}, \bibinfo {author} {\bibfnamefont
  {J.}~\bibnamefont {Izaac}}, \bibinfo {author} {\bibfnamefont
  {Z.}~\bibnamefont {Jiang}}, \bibinfo {author} {\bibfnamefont
  {X.}~\bibnamefont {Liu}}, \bibinfo {author} {\bibfnamefont {S.}~\bibnamefont
  {McArdle}}, \bibinfo {author} {\bibfnamefont {M.}~\bibnamefont {Neeley}},
  \bibinfo {author} {\bibfnamefont {T.}~\bibnamefont {O’Brien}}, \bibinfo
  {author} {\bibfnamefont {B.}~\bibnamefont {O’Gorman}}, \bibinfo {author}
  {\bibfnamefont {I.}~\bibnamefont {Ozfidan}}, \bibinfo {author} {\bibfnamefont
  {M.~D.}\ \bibnamefont {Radin}}, \bibinfo {author} {\bibfnamefont
  {J.}~\bibnamefont {Romero}}, \bibinfo {author} {\bibfnamefont {N.~P.~D.}\
  \bibnamefont {Sawaya}}, \bibinfo {author} {\bibfnamefont {B.}~\bibnamefont
  {Senjean}}, \bibinfo {author} {\bibfnamefont {K.}~\bibnamefont {Setia}},
  \bibinfo {author} {\bibfnamefont {S.}~\bibnamefont {Sim}}, \bibinfo {author}
  {\bibfnamefont {D.~S.}\ \bibnamefont {Steiger}}, \bibinfo {author}
  {\bibfnamefont {M.}~\bibnamefont {Steudtner}}, \bibinfo {author}
  {\bibfnamefont {Q.}~\bibnamefont {Sun}}, \bibinfo {author} {\bibfnamefont
  {W.}~\bibnamefont {Sun}}, \bibinfo {author} {\bibfnamefont {D.}~\bibnamefont
  {Wang}}, \bibinfo {author} {\bibfnamefont {F.}~\bibnamefont {Zhang}},\ and\
  \bibinfo {author} {\bibfnamefont {R.}~\bibnamefont {Babbush}},\ }\bibfield
  {title} {\bibinfo {title} {Openfermion: the electronic structure package for
  quantum computers},\ }\href {https://doi.org/10.1088/2058-9565/ab8ebc}
  {\bibfield  {journal} {\bibinfo  {journal} {Quantum Science and Technology}\
  }\textbf {\bibinfo {volume} {5}},\ \bibinfo {pages} {034014} (\bibinfo {year}
  {2020})}\BibitemShut {NoStop}%
\bibitem [{\citenamefont {Suzuki}\ \emph {et~al.}(2021)\citenamefont {Suzuki},
  \citenamefont {Kawase}, \citenamefont {Masumura}, \citenamefont {Hiraga},
  \citenamefont {Nakadai}, \citenamefont {Chen}, \citenamefont {Nakanishi},
  \citenamefont {Mitarai}, \citenamefont {Imai}, \citenamefont {Tamiya},
  \citenamefont {Yamamoto}, \citenamefont {Yan}, \citenamefont {Kawakubo},
  \citenamefont {Nakagawa}, \citenamefont {Ibe}, \citenamefont {Zhang},
  \citenamefont {Yamashita}, \citenamefont {Yoshimura}, \citenamefont
  {Hayashi},\ and\ \citenamefont {Fujii}}]{Suzuki2021}%
  \BibitemOpen
  \bibfield  {author} {\bibinfo {author} {\bibfnamefont {Y.}~\bibnamefont
  {Suzuki}}, \bibinfo {author} {\bibfnamefont {Y.}~\bibnamefont {Kawase}},
  \bibinfo {author} {\bibfnamefont {Y.}~\bibnamefont {Masumura}}, \bibinfo
  {author} {\bibfnamefont {Y.}~\bibnamefont {Hiraga}}, \bibinfo {author}
  {\bibfnamefont {M.}~\bibnamefont {Nakadai}}, \bibinfo {author} {\bibfnamefont
  {J.}~\bibnamefont {Chen}}, \bibinfo {author} {\bibfnamefont {K.~M.}\
  \bibnamefont {Nakanishi}}, \bibinfo {author} {\bibfnamefont {K.}~\bibnamefont
  {Mitarai}}, \bibinfo {author} {\bibfnamefont {R.}~\bibnamefont {Imai}},
  \bibinfo {author} {\bibfnamefont {S.}~\bibnamefont {Tamiya}}, \bibinfo
  {author} {\bibfnamefont {T.}~\bibnamefont {Yamamoto}}, \bibinfo {author}
  {\bibfnamefont {T.}~\bibnamefont {Yan}}, \bibinfo {author} {\bibfnamefont
  {T.}~\bibnamefont {Kawakubo}}, \bibinfo {author} {\bibfnamefont {Y.~O.}\
  \bibnamefont {Nakagawa}}, \bibinfo {author} {\bibfnamefont {Y.}~\bibnamefont
  {Ibe}}, \bibinfo {author} {\bibfnamefont {Y.}~\bibnamefont {Zhang}}, \bibinfo
  {author} {\bibfnamefont {H.}~\bibnamefont {Yamashita}}, \bibinfo {author}
  {\bibfnamefont {H.}~\bibnamefont {Yoshimura}}, \bibinfo {author}
  {\bibfnamefont {A.}~\bibnamefont {Hayashi}},\ and\ \bibinfo {author}
  {\bibfnamefont {K.}~\bibnamefont {Fujii}},\ }\bibfield  {title} {\bibinfo
  {title} {Qulacs: a fast and versatile quantum circuit simulator for research
  purpose},\ }\href {https://doi.org/10.22331/q-2021-10-06-559} {\bibfield
  {journal} {\bibinfo  {journal} {{Quantum}}\ }\textbf {\bibinfo {volume}
  {5}},\ \bibinfo {pages} {559} (\bibinfo {year} {2021})}\BibitemShut {NoStop}%
\bibitem [{\citenamefont {Anand}\ \emph {et~al.}(2022)\citenamefont {Anand},
  \citenamefont {Schleich}, \citenamefont {Alperin-Lea}, \citenamefont
  {Jensen}, \citenamefont {Sim}, \citenamefont {Díaz-Tinoco}, \citenamefont
  {Kottmann}, \citenamefont {Degroote}, \citenamefont {Izmaylov},\ and\
  \citenamefont {Aspuru-Guzik}}]{Anand2022}%
  \BibitemOpen
  \bibfield  {author} {\bibinfo {author} {\bibfnamefont {A.}~\bibnamefont
  {Anand}}, \bibinfo {author} {\bibfnamefont {P.}~\bibnamefont {Schleich}},
  \bibinfo {author} {\bibfnamefont {S.}~\bibnamefont {Alperin-Lea}}, \bibinfo
  {author} {\bibfnamefont {P.~W.~K.}\ \bibnamefont {Jensen}}, \bibinfo {author}
  {\bibfnamefont {S.}~\bibnamefont {Sim}}, \bibinfo {author} {\bibfnamefont
  {M.}~\bibnamefont {Díaz-Tinoco}}, \bibinfo {author} {\bibfnamefont {J.~S.}\
  \bibnamefont {Kottmann}}, \bibinfo {author} {\bibfnamefont {M.}~\bibnamefont
  {Degroote}}, \bibinfo {author} {\bibfnamefont {A.~F.}\ \bibnamefont
  {Izmaylov}},\ and\ \bibinfo {author} {\bibfnamefont {A.}~\bibnamefont
  {Aspuru-Guzik}},\ }\bibfield  {title} {\bibinfo {title} {A quantum computing
  view on unitary coupled cluster theory},\ }\href
  {https://doi.org/10.1039/D1CS00932J} {\bibfield  {journal} {\bibinfo
  {journal} {Chem. Soc. Rev.}\ }\textbf {\bibinfo {volume} {51}},\ \bibinfo
  {pages} {1659} (\bibinfo {year} {2022})}\BibitemShut {NoStop}%
\bibitem [{qur(2022)}]{quri-parts}%
  \BibitemOpen
  \href {https://github.com/QunaSys/quri-parts} {\bibinfo {title} {{QURI
  Parts}}} (\bibinfo {year} {2022}),\ \bibinfo {note}
  {\url{https://github.com/QunaSys/quri-parts}}\BibitemShut {NoStop}%
\bibitem [{\citenamefont {Hirsbrunner}\ \emph {et~al.}(2024)\citenamefont
  {Hirsbrunner}, \citenamefont {Chamaki}, \citenamefont {Mullinax},\ and\
  \citenamefont {Tubman}}]{Hirsbrunner2024beyondmp}%
  \BibitemOpen
  \bibfield  {author} {\bibinfo {author} {\bibfnamefont {M.~R.}\ \bibnamefont
  {Hirsbrunner}}, \bibinfo {author} {\bibfnamefont {D.}~\bibnamefont
  {Chamaki}}, \bibinfo {author} {\bibfnamefont {J.~W.}\ \bibnamefont
  {Mullinax}},\ and\ \bibinfo {author} {\bibfnamefont {N.~M.}\ \bibnamefont
  {Tubman}},\ }\bibfield  {title} {\bibinfo {title} {Beyond {MP}2
  initialization for unitary coupled cluster quantum circuits},\ }\href
  {https://doi.org/10.22331/q-2024-11-26-1538} {\bibfield  {journal} {\bibinfo
  {journal} {{Quantum}}\ }\textbf {\bibinfo {volume} {8}},\ \bibinfo {pages}
  {1538} (\bibinfo {year} {2024})}\BibitemShut {NoStop}%
\bibitem [{\citenamefont {Matsuzawa}\ and\ \citenamefont
  {Kurashige}(2020)}]{Matsuzawa2020}%
  \BibitemOpen
  \bibfield  {author} {\bibinfo {author} {\bibfnamefont {Y.}~\bibnamefont
  {Matsuzawa}}\ and\ \bibinfo {author} {\bibfnamefont {Y.}~\bibnamefont
  {Kurashige}},\ }\bibfield  {title} {\bibinfo {title} {Jastrow-type
  decomposition in quantum chemistry for low-depth quantum circuits},\ }\href
  {https://doi.org/10.1021/acs.jctc.9b00963} {\bibfield  {journal} {\bibinfo
  {journal} {Journal of Chemical Theory and Computation}\ }\textbf {\bibinfo
  {volume} {16}},\ \bibinfo {pages} {944} (\bibinfo {year} {2020})}\BibitemShut
  {NoStop}%
\bibitem [{\citenamefont {Motta}\ \emph {et~al.}(2023)\citenamefont {Motta},
  \citenamefont {Sung}, \citenamefont {Whaley}, \citenamefont {Head-Gordon},\
  and\ \citenamefont {Shee}}]{motta2023Bridging}%
  \BibitemOpen
  \bibfield  {author} {\bibinfo {author} {\bibfnamefont {M.}~\bibnamefont
  {Motta}}, \bibinfo {author} {\bibfnamefont {K.~J.}\ \bibnamefont {Sung}},
  \bibinfo {author} {\bibfnamefont {K.~B.}\ \bibnamefont {Whaley}}, \bibinfo
  {author} {\bibfnamefont {M.}~\bibnamefont {Head-Gordon}},\ and\ \bibinfo
  {author} {\bibfnamefont {J.}~\bibnamefont {Shee}},\ }\bibfield  {title}
  {\bibinfo {title} {Bridging physical intuition and hardware efficiency for
  correlated electronic states: the local unitary cluster jastrow ansatz for
  electronic structure},\ }\href {https://doi.org/10.1039/D3SC02516K}
  {\bibfield  {journal} {\bibinfo  {journal} {Chem. Sci.}\ }\textbf {\bibinfo
  {volume} {14}},\ \bibinfo {pages} {11213} (\bibinfo {year}
  {2023})}\BibitemShut {NoStop}%
\bibitem [{\citenamefont {Akahoshi}\ \emph {et~al.}(2024)\citenamefont
  {Akahoshi}, \citenamefont {Maruyama}, \citenamefont {Oshima}, \citenamefont
  {Sato},\ and\ \citenamefont {Fujii}}]{Akahoshi2024}%
  \BibitemOpen
  \bibfield  {author} {\bibinfo {author} {\bibfnamefont {Y.}~\bibnamefont
  {Akahoshi}}, \bibinfo {author} {\bibfnamefont {K.}~\bibnamefont {Maruyama}},
  \bibinfo {author} {\bibfnamefont {H.}~\bibnamefont {Oshima}}, \bibinfo
  {author} {\bibfnamefont {S.}~\bibnamefont {Sato}},\ and\ \bibinfo {author}
  {\bibfnamefont {K.}~\bibnamefont {Fujii}},\ }\bibfield  {title} {\bibinfo
  {title} {Partially fault-tolerant quantum computing architecture with
  error-corrected clifford gates and space-time efficient analog rotations},\
  }\href {https://doi.org/10.1103/PRXQuantum.5.010337} {\bibfield  {journal}
  {\bibinfo  {journal} {PRX Quantum}\ }\textbf {\bibinfo {volume} {5}},\
  \bibinfo {pages} {010337} (\bibinfo {year} {2024})}\BibitemShut {NoStop}%
\bibitem [{\citenamefont {Campbell}(2019)}]{campbel2019}%
  \BibitemOpen
  \bibfield  {author} {\bibinfo {author} {\bibfnamefont {E.}~\bibnamefont
  {Campbell}},\ }\bibfield  {title} {\bibinfo {title} {Random compiler for fast
  hamiltonian simulation},\ }\href
  {https://doi.org/10.1103/PhysRevLett.123.070503} {\bibfield  {journal}
  {\bibinfo  {journal} {Phys. Rev. Lett.}\ }\textbf {\bibinfo {volume} {123}},\
  \bibinfo {pages} {070503} (\bibinfo {year} {2019})}\BibitemShut {NoStop}%
\end{thebibliography}%
%%% --------------- %%%

%%% --------------- %%%
\appendix
\section{Series expansion for time-dependent probability $P_\mu(t)$}
\label{app:seriesexpansion}
We present the derivation of Eq.~\eqref{eq:probabilityschematicseriesexpansion} in the main text.

We consider the series expansion $e^{z} =\sum_{k=0}^{\infty}\frac{z^k}{k!}$, which leads to 
\begin{align}
P_\mu(t) &= \langle \mu | e^{-i\hat{H}t}|\psi_I \rangle \langle \psi_I | e^{i\hat{H}t}|\mu \rangle=\sum_{k_1=0, k_2=0}^{\infty}(-i)^{k_1} (i)^{ k_2} t^{k_1+k_2} \frac{ \langle  \mu | \hat{H}^{k_1} | \psi_{I} \rangle \langle \psi_{I} | \hat{H}^{k_2} | \mu\rangle }{k_1! k_2!}.
\label{eq:probabilityseriesexpansionintermediatestep}
\end{align}
Let us consider the terms where $k_1=0$ or $k_2=0$ separately.
Defining $b_{\mu}= \langle \mu| \psi_I \rangle$ and considering $k_1=0$ or $k_2=0$ gives $ (i)^{k_2} b_{\mu} \langle | \psi_{I} | \hat{H}^{k_2} |\mu \rangle $ and $ (-i)^{k_1} (b_{\mu})^*  \mel{\mu}{\hat{H}^{k_1}}{\psi_{I}}$, respectively, so we can gather these two terms in one by taking 2 times the real value.
The special case of $k_1=k_2=0$ leads to $|b_{\mu}|^2$, so overall this part of the double sum can be expressed as 
\begin{align}
&|b_{\mu}|^2 +2 \sum_{k=1}^{\infty} t^{k} \frac{ \text{Re}\left[ (i)^{k} b_{\mu} \langle | \psi_{I} | \hat{H}^{k} |\mu \rangle \right] }{k!}.
\end{align}
The remaining terms are 
\begin{align}
\sum_{k_1=1, k_2=1}^{\infty}(-i)^{k_1}(i)^{k_2} t^{k_1+k_2} \frac{ \langle  \mu | \hat{H}^{k_1} | \psi_{I} \rangle \langle \psi_{I} | \hat{H}^{k_2} | \mu\rangle }{k_1! k_2!}.
\end{align}
We then consider the case where $k_1=k_2$ and $k_1 \neq k_2$ separately. For $k_1=k_2$, the sum straightforwardly gives 
\begin{align}
\sum_{k=1}^{\infty} t^{2k} \frac{ |\langle  \mu | \hat{H}^{k} | \psi_{I} \rangle |^2}{(k!)^2}
\end{align}
For the latter, we can rewrite the sum as 2 times the real part, restricting the sum to values $k_2>k_1$, which gives 
\begin{align}
\sum_{k_1=1,k_2=1, k_2>k_1}^{\infty} 2 t^{k_1+k_2} \frac{  \text{Re}\left[(-i)^{k_1}(i)^{k_2} \langle  \mu | \hat{H}^{k_1} | \psi_{I} \rangle \langle \psi_{I} | \hat{H}^{k_2} | \mu\rangle\right] }{k_1! k_2!}
\end{align}
This sum can be given a more useful schematic form by substituting $k_2$ with the index $k_2=k_1+k'$ where $k'=1,2...,$, leading to
\begin{align}
\sum_{k_1=1,k'=1}^{\infty} 2 t^{2 k_1+k'} \frac{  \text{Re}\left[(i)^{k'} \langle  \mu | \hat{H}^{k_1} | \psi_{I} \rangle \langle \psi_{I} | \hat{H}^{k_1+k'} | \mu\rangle\right] }{k_1! (k_1+k')!}
\end{align}
We can then relabel all $k_1$ as $k$, which leads to Eq.~\eqref{eq:probabilityschematicseriesexpansion} in the main text when putting all terms together.

%%% --------------- %%%
\section{Supplementary numerical results}

\subsection{Dependence of TE-QSCI on the input state and bond distance for 16-qubit hydrogen chain}
\label{app:fidelitydependence}
\begin{figure*}[htb!]
\includegraphics[width=1\linewidth]{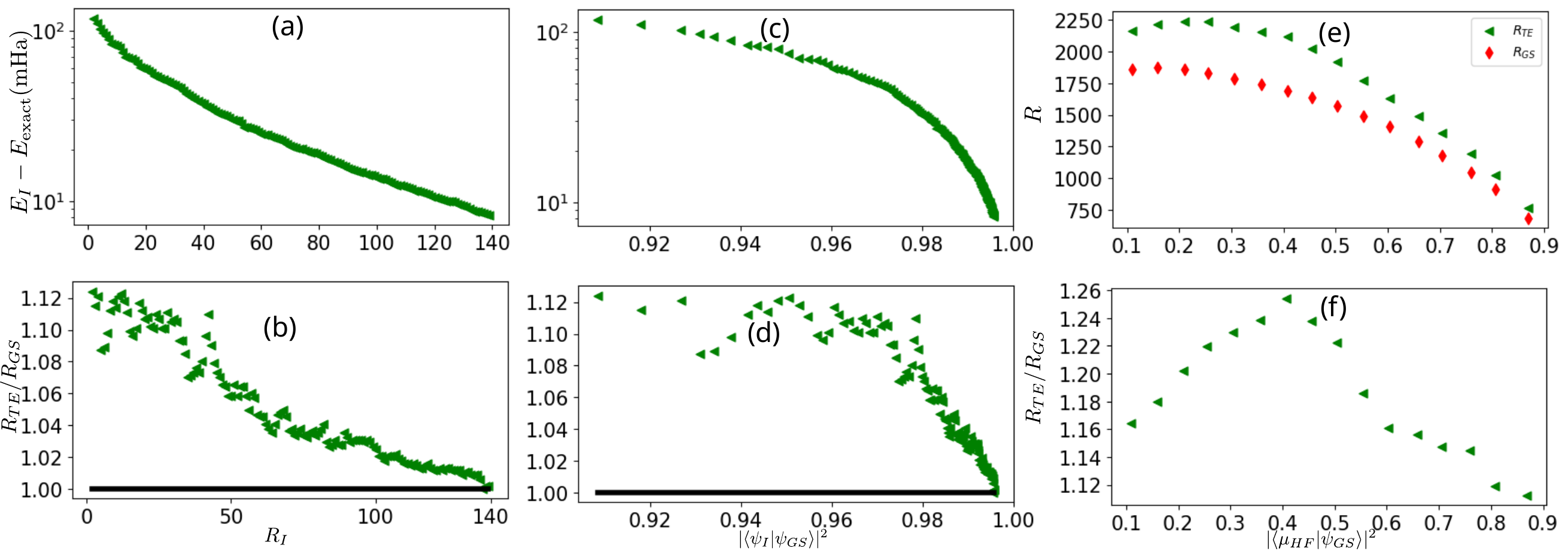}
\caption{Dependence of the subspace dimension $R_{TE}$ in single-time TE-QSCI on the fidelity between the initial state of the time evolution and the exact ground state studied for 16-qubit hydrogen chain \ce{H8}.
(a)-(d): The initial state is prepared by the method described in the main text with fixing the Hamiltonian.
$R_I$ is the dimension of the subspaces used for preparing the initial state.
Panels (a) and (c) show the energy of the input state as a function of $R_I$ and the fidelity $\abs{\braket{\psi_{I}}{\psi_{GS}}}^2$, respectively, while panels (b) and (d) show $R_{TE}/R_{GS}$ (the ``optimal" time is always $t=0.75$ as this is the case for a majority of the input states).
(e), (f): The initial state is the Hartree-Fock state but the Hamiltonian is varied by changing the atomic distance of the hydrogen chain from $1.0$~\AA{} to $2.5$~\AA.
We plot the dimensions of the subspaces $R_{TE}$ (the optimal time is $t=1.4$) for single-time QSCI and $R_{GS}$ for GS-QSCI in panel (e) and $R_{TE}/R_{GS}$ in panel (f) as a function of the fidelity.}
\label{fig:fidelitydependentqsci} 
\end{figure*}

In this section, we study the accuracy dependence on the fidelity between the initial state of the time evolution of single-time TE-QSCI and the exact ground state, using the 16 qubit hydrogen chain \ce{H8} as an example.
We systematically change the fidelity of the initial state in two ways.
The first one uses an artificial input state obtained by classical calculation while the Hamiltonian is fixed.
The second one, on the contrary, keeps the initial state as the Hartree-Fock state and varies the Hamiltonian.
We stress that the fidelity between the time-evolved state and the exact ground state is constant during the time evolution, so the initial state of the time evolution solely determines the fidelity.
We denote the smallest dimension of the subspace required to obtain $10^{-3}$ Hartree accuracy for single-time TE-QSCI at the optimal time and GS-QSCI as $R_{TE}$ and $R_{GS}$, respectively.

Let us explain the first approach and its result.
We prepare the initial state of the time evolution by choosing $R_I (< R_{GS}=685)$ computational basis states (electron configurations) using the amplitudes of the exact ground state, sorted in descending order.
Classical exact diagonalization in the $R_I$-dimensional subspace is performed and the resulting lowest-energy eigenstate is employed as the input for the time evolution.
We can vary the fidelity of the initial state by changing the value of $R_I$.
Figure~\ref{fig:fidelitydependentqsci}(a)-(d) shows the lowest energy $E_I$ in the $R_I$-dimensional subspace and $R_{TE}/R_{GS}$.
As expected, when $R_I$ becomes large and the fidelity between the initial state and the exact ground state gets close to unity, the ratio $R_{TE}/R_{GS}$ becomes smaller.
Note that even for the data points close to the fidelity~$\sim 1$, the value of $R_I \sim 140$ is much smaller than $R_{GS}=685$, leading to $E_I-E_{\text{exact}}\sim 10^{-2}$ Hartree.
This means the value of $R_{TE}/R_{GS} \sim 1$ indicating that results are on par with GS-QSCI is due to TE-QSCI generating important configurations not present in the initial state.

The second approach we take to vary the fidelity of the initial state is increasing the bond length (atomic distance) of \ce{H8} from the original value of $1$~\AA.
Because the Hartree-Fock state becomes a worse approximation of the ground state for larger bond lengths, the fidelity correspondingly decreases.
In Fig.~\ref{fig:fidelitydependentqsci}(e,f) we plot $R_{TE}$, $R_{GS}$ and $R_{TE}/R_{GS}$, respectively, for different bond lengths between $1$~\AA{} and $2.5$~\AA{} as a function of $\abs{\braket{\mu_{HF}}{\psi_{GS}}}^2$, where $\ket{\mu_{HF}}$ ($\ket{\psi_{GS}}$) is the Hartree-Fock state (exact ground state).
Interestingly, in this case the ratio $R_{TE}/R_{GS}$ is not monotonically decreasing with the fidelity.
It is also surprising that single-time QSCI can give an accurate energy with relatively small additional classical computational resources ($R_{TE} \lesssim 1.25 R_{GS}$) despite the small fidelity.

In short, the results in this section suggest that while a large fidelity between the initial state of the time evolution and the exact ground state saves classical computational cost (small $R_{TE}/R_{GS)}$), it is not the sole factor determining the performance of single-time TE-QSCI. 

\subsection{Fidelity and the energy conservation of the Trotterized time evolution}
\label{app:trottererror}
\begin{figure*}[htb]
\includegraphics[width=1\linewidth]{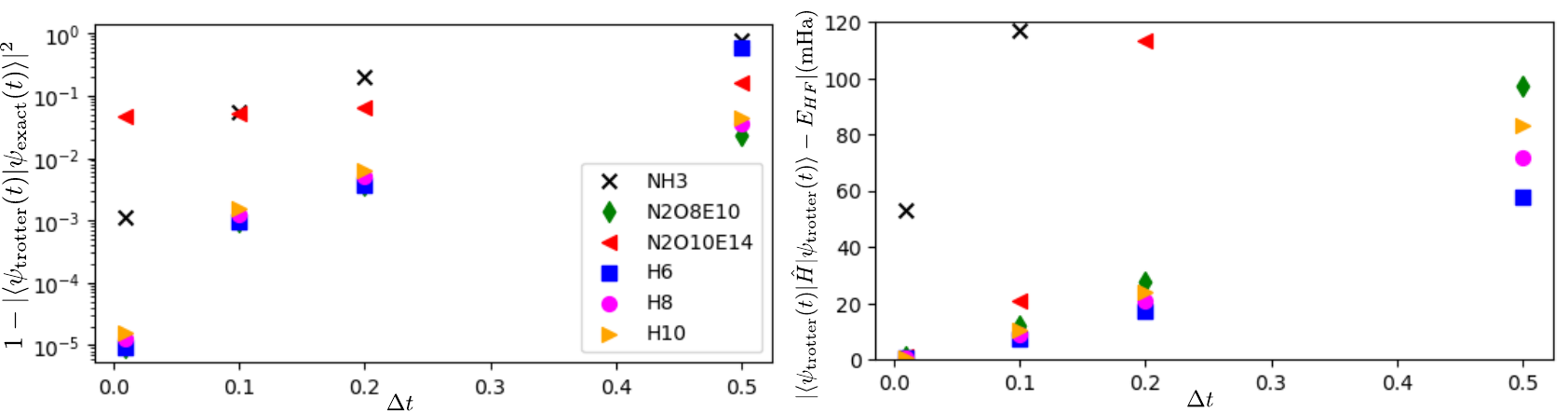}
\caption{
(a) Infidelity between the time-evolved state approximated by Trotterization and the exact time-evolved state, $1-|\braket{\psi_{\text{trotter}}(t)}{\psi_{\text{exact}}(t)}|^2$. 
(b) Absolute difference between the expectation value of the Hamiltonian and the initial energy, $|\ev{\hat{H}}{\psi_{\text{trotter}}(t)}-E_{HF}|$ as a function of the Trotter step size $\Delta t$ for specific choices of the time, corresponding to those of Fig.~\ref{fig:Rdependentqsci}.
}
\label{fig:timedependenttrottererror} 
\end{figure*}

In Sec.~\ref{subsec:performance of single TE QSCI}, it was seen that single-time TE-QSCI is relatively robust towards the Trotter error for hydrogen chains, while it is somewhat less robust for \ce{N2} and \ce{NH3}.
In this section, we present more detailed data of the error associated with the Trotterized time evolution, finding the same overall trend.
In Fig.~\ref{fig:timedependenttrottererror}, we plot the infidelity between the state after the Trotterized time-evolution and the exact time-evolved state, $1-|\langle \psi_{\text{trotter}}(t)|\psi_{\text{exact}}(t) \rangle|^2$ as well as the violation of the energy conservation during the time evolution $|\langle\psi_{\text{trotter}}(t)|\hat{H}| \psi_{\text{trotter}}(t) \rangle-E_{HF} |$, at the optimal times corresponding to Fig.~\ref{fig:Rdependentqsci}.
An observed overall tendency is that the hydrogen chains exhibit smaller values both for the infidelity and the violation of the energy conservation than \ce{N2} and \ce{NH3}, which may result in the accurate output energy of single-time TE-QSCI seen in Fig.~\ref{fig:timedependentqsci}.
Although it is hard to make any strong conclusions based on these results, the accuracy of single-time TE-QSCI generally get worse if both the infidelity and energy error are large.
Moreover, it is interesting and promising that single-time TE-QSCI can give the accurate energy error of $10^{-3}$ Hartree even when the Trotter error measured by the violation of the energy conservation is much larger than it.

\subsection{Time evolution for the UCCSD initial state}
\label{app:UCCSDtimeevolution}

\begin{figure*}[htb]
\includegraphics[width=1\linewidth]{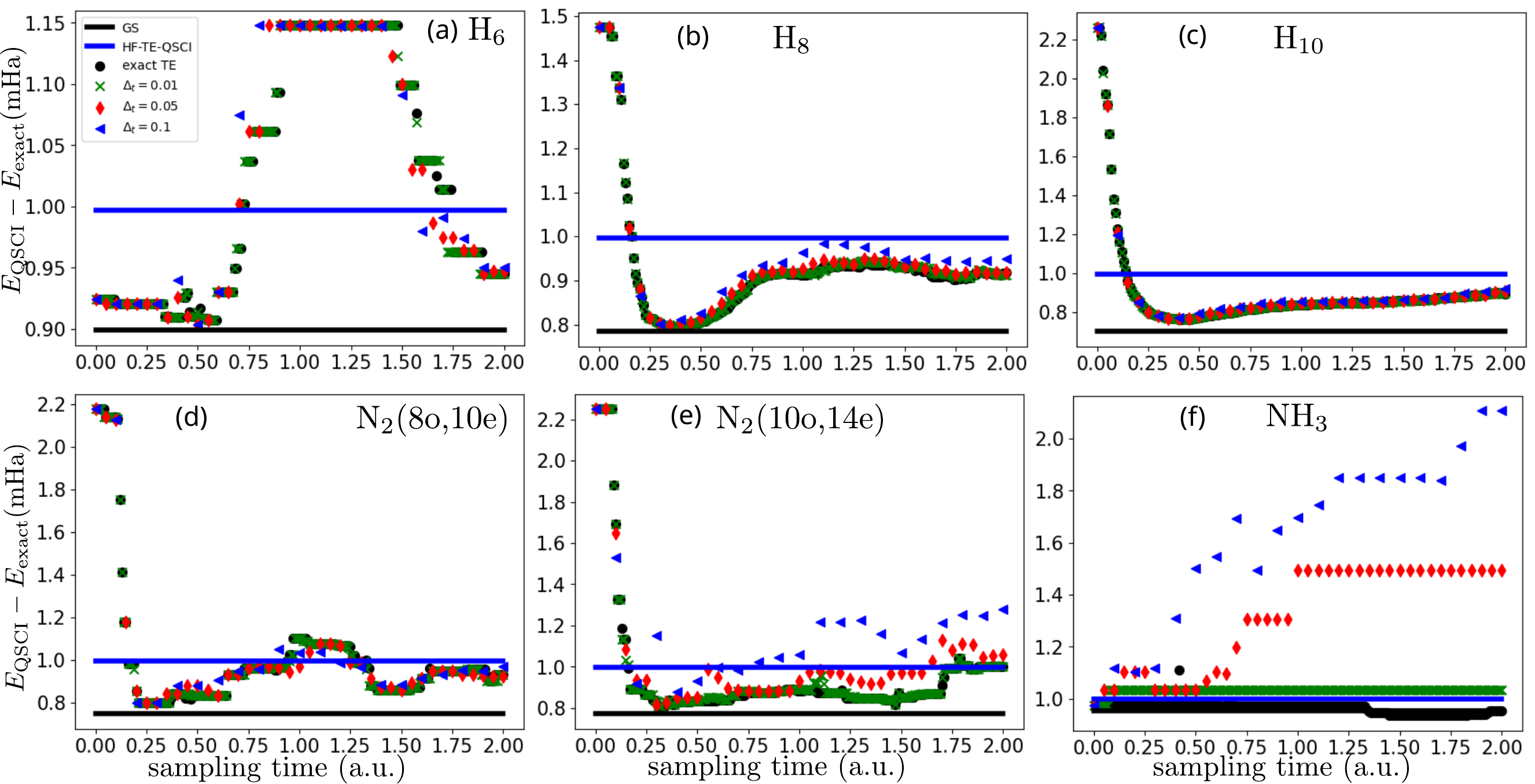}
\caption{Difference between the exact ground-state energy and the energy obtained by single-time QSCI taking an initial state as the UCCSD ansatz with CCSD amplitudes (UCCSD-TE-QSCI).
Various Trotter step sizes $\Delta t$ are taken for the approximation of the time-evolution operator.
The solid black line corresponds to GS-QSCI, while the solid blue line corresponds to the energy obtained from HF-TE-QSCI at the optimal times used in Fig.~\ref{fig:Rdependentqsci}.
The black circles represent the results of the exact time evolution.
The dimensions of the subspace in TE-QSCI are $R=87, 762, 5780, 123, 141$ and $78$ for panels (a)-(f), respectively.}
\label{fig:timedependentqsciUCCSD} 
\end{figure*}

In Sec.~\ref{subsec:comparison with CCSD} we found that single-time TE-QSCI with an initial state defined by the UCCSD ansatz with classically calculated CCSD amplitudes using a single Trotter step (UCCSD-TE-QSCI) can lead to better results than UCCSD-QSCI and single-time TE-QSCI with the Hartree-Fock initial state (HF-TE-QSCI).
Here, we supplement the results in the main text by showing the time dependence of the obtained energy of UCCSD-TE-QSCI using various sizes of Trotterization for the time-evolution operator, similarly to Fig.~\ref{fig:timedependentqsci}. The values of $R$, or the dimension of the subspace of QSCI, are chosen to be the smallest $R$ that gives an accuracy of $10^{-3}$ Hartree for HF-TE-QSCI with the exact time evolution at the optimal time.
The results are shown in Fig.~\ref{fig:timedependentqsciUCCSD}.
For all molecules investigated, UCCSD-TE-QSCI with some optimal time ranges can yield the same or better energies than the exact time-evolved HF-TE-QSCI at the optimal time.
As for the comparison to UCCSD-QSCI (corresponding to the values at $t=0$),
we see a drastic improvement of energies for \ce{H8},\ce{H10} and \ce{N2}, while it is not obvious for \ce{NH3} and \ce{H6}, where the classical CCSD calculation was already quite accurate (see Table~\ref{tab:UCCSDcomparison}).
Interestingly, the optimal times for UCCSD-TE-QSCI are smaller than those for HF-TE-QSCI. If this property holds in general, it suggests that the UCCSD initial state could potentially decrease the number of Trotter steps and the number of quantum gates for the time-evolution operator.

\end{document}